\documentclass[reprint,twocolumn,showpacs]{revtex4-1}

\usepackage{amsfonts,amssymb,amsmath,amsthm,todo,multirow,}
\usepackage{mathtools} 
\usepackage{textcomp} 
\usepackage{graphicx} 
\usepackage{bm}
\usepackage{tikz}
\usetikzlibrary{shapes,arrows,calc,shapes.misc,backgrounds,
decorations.pathreplacing,positioning}
\usepackage{dcolumn}
\usepackage{epstopdf}

\newcommand{\ket}[1]{\left| #1 \right>} 
\newcommand{\bra}[1]{\left< #1 \right|} 

\newcommand{\N}{N \setminus (O,t)}
\newcommand{\Ot}{O \setminus t}
\newcommand{\bNu}{\vec{b},\nu_{c}}

\newtheorem{theorem}{Theorem}
\newtheorem{defn}[theorem]{Definition}

\newtheorem{lemma}[theorem]{Lemma}

\newenvironment{definition}[1][Definition]{\begin{trivlist}
\item[\hskip \labelsep {\bfseries #1}]}{\end{trivlist}}

\begin{document}

\preprint{APS/123-QED}


\title{Quantum protocols within Spekkens' toy model}

\author{Leonardo Disilvestro} \email{disilves@telecom-paristech.fr}
\author{Damian Markham}%
 
\affiliation{%
LTCI, CNRS, Telecom ParisTech, Univ Paris-Saclay, 75013 Paris, France
}%
\begin{abstract}
Quantum mechanics is known to provide significant improvements in information processing tasks when compared to classical models.
 These advantages range from computational speeds-up to security improvements. 
 A key question is where these advantages come from. 
 The toy model developed by Spekkens [R. W. Spekkens PRA \textbf{75}, 032110 (2012)]  mimics many of the features of quantum mechanics, such as entanglement and no-cloning, regarded as being important in this regard, despite being a local hidden variable theory.
In this work we  study several protocols  within Spekkens' toy model where we see it can also mimic the advantages and limitations shown in the quantum case. 
 We first provide explicit proofs for the impossibility of toy bit-commitment and the existence of a toy error correction protocol and consequent $k-$threshold secret sharing. 
 Then, defining a toy computational model based on the quantum one-way computer we prove the existence of blind and verified protocols. Importantly, these two last quantum protocols are known to achieve a better-than-classical security. 
 Our results suggest that such quantum improvements need not arise from any Bell-type non locality or contextuality, but rather as a consequence of steering correlations.
\end{abstract}

\maketitle

\section{Introduction and motivation}

Despite the undeniable ability of quantum theory to accurately describe the behaviour of atomic and subatomic systems, there exist different views on what a quantum state actually is. Given a system of interest the \textit{epistemic} view takes a quantum state to be a \textit{description} of an agent's knowledge about the system, while the \textit{ontic} view takes it to be a \textit{property} of the system. In~\cite{Spekkens2007} Spekkens introduced an epistemic local hidden variable (LHV) theory, the toy model, in support of the epistemic view. This toy model can be seen as fundamentally arising as an epistemic restriction of a statistical theory associated to classical bits~\cite{Spekkens2016}, and is phenomenologically very close to a subset of quantum mechanics know as the stabilizer formalism for qubits~\cite{Coecke2011,Pusey2012}. In particular it reproduces many aspects which were often considered as `signatures' of quantum behaviors. A non exhaustive list of aspects includes entanglement, steering, interference, incompatibility of measurements, no-cloning, and teleportation~\cite{Spekkens2007}. The fundamental insight here, is that these features cannot be taken as `signatures' of quantum theory. However it should also be pointed out that the local aspect of the theory implies no violations of Bell inequalities nor any form of contexuality is ever witnessed in the toy model~\cite{Spekkens2007}. 

In this work we pursue Spekkens' original interest by considering which quantum protocols can be run in the toy model. 
To do so we approach protocols using the stabilizer formalism - an efficient way of representing families of states and operators which is useful across much of quantum information.
We study whether stabilizer based quantum protocols can be meaningfully defined or replicated in the toy model. We achieve this by extending the toy-stabilizer formalism of~\cite{Pusey2012}, and by defining a toy model of computation similar to the measurement based one in quantum theory~\cite{Raussendorf2001}, hence allowing a wide range of quantum protocols from error correction~\cite{Gottesman1997} to blind and verified computations~\cite{Broadbent2008,Fitzsimons2012} to be analyzed in the toy model.

\bigskip

In the toy model the underlying physical states are represented by classical variables. These are the \textit{ontic states} of the theory, and for a single toy system they correspond to four possible configurations in phase-space~\cite{Spekkens2007,Coecke2011}. However these ontic variables are not accessible to an observer, who is forced to describe the underlying ontic reality statistically. These statistical descriptions of the ontic states define the \textit{epistemic states} of the toy theory. The peculiarity of the toy model is then the imposition of a \textit{restriction} on what an observer is allowed to know. The imposition goes under the name of `\textit{knowledge balance principle}' (KBP), the idea being that the amount one can know about a system is at best matched by ones ignorance. While in its generality its result is somewhat awkward to state~\cite{Coecke2011}, for a single toy system it says that the best description an observer can give is that the system is in one of two ontic states, with equal probability~\cite{Spekkens2007,Pusey2012}. Crucial to Spekkens' argument is the fact that the epistemic states, and not the ontic ones, resemble qubits.

There are, however, two key places where the toy theory fails to reproduce quantum behaviours: it is impossible to violate Bell's type inequalities or show any form of contextuality~\cite{Spekkens2007,Spekkens2016,Pusey2012,Coecke2011}. Where contextuality is broadly defined as the impossibility of simultaneously assigning a definite measurement outcome to all observables of a given system~\cite{Kochen1967}. This is in sharp contrast with quantum theory which is known to be Bell non-local and contextual.

There are two further subtle points about the toy model that are prudent at this point. \textit{Firstly}, despite the toy model resembling the stabilizer framework of quantum theory, it is \textit{not} a subset of quantum mechanics: it is a genuinely different theory as states and transformations are not in a one-to-one correspondence to those of quantum stabilizers~\cite{Spekkens2007,Pusey2012,Coecke2011}. And \textit{secondly}, the toy model is computationally very weak: a toy computer can at most compute linear functions and it efficiently simulable by a classical computer~\cite{Spekkens2007,Pusey2012}.

Indeed our interest in the toy protocols we introduce in section~\eqref{sec:Results} is not to lead the way to some useful `toy implementation', nor support the toy theory as a physical theory, but rather as way to better understand whether Bell non-locality or contextuality lie behind the respective quantum protocols: the local nature of the toy model allows for a natural test theory to study whether these properties are truly necessary for the quantum protocols.

From a more practical perspective we also recall that the study of epistemically restricted classical theories also has a very rooted physical grounding~\cite{Spekkens2016}. If one considers the classical statistical theory of optics and introduces an epistemic restriction in the form of an uncertainty principle, Gaussian quantum optics is obtained~\cite{Bartlett2012}. This is a very well known example of a `toy theory' which supports protocols showing quantum advantage~\cite{GaussianQuantumInformation}, and indeed observes some common limitations for example  the results in~\cite{Leverrier2014} where a no-go theorem for Gaussian quantum bit commitment is obtained.

In this work we have chosen the toy stabilizer formalism of~\cite{Pusey2012} as it is the most suited  to talk about the protocols we study, but this is not the only alternative to the original notation presented in~\cite{Spekkens2007}. In particular, there exists another framework based on the categorical approach to quantum theory~\cite{Coecke2011} which has recently led to the formulation of an universal, sound, and complete graphical language for the toy mode known as the ZX-calculusl~\cite{Backens2015}. 

The paper will develop as follows. In section \eqref{sec:Background} the toy model is formally introduced. The ideas and notation presented in~\cite{Spekkens2007,Pusey2012} are reviewed and some new concepts and extensions are defined in order to study protocols. 
In particular the new additions include the introduction of a partial trace operation, explicit toy purification, and irreversible transformations. In section~\eqref{sec:Results} our main results are introduced. 
We start in~\eqref{sec:Recovering_Commutation_Relations} and~\eqref{sec:Encoding_Information} by operationally recovering the quantum-like anticommutation relations between toy transformations and toy measurements and we then proceed to define how to encode information within a toy stabilizer. ~\eqref{sec:Bit_Commitment_No_Delition_Theorem} shows a no-go theorem, as conjectured by Spekkens in~\cite{Spekkens2007}, for perfect and imperfect toy bit commitments schemes and provides a toy no-deletion theorem, then in~\eqref{sec:Error_Correction} it is proved that toy error correction and toy secret sharing protocols are possible within the toy model whenever they are possible in the quantum case. In section~\eqref{sec:MBTC}, a definition of measurement based toy computation model followed in~\eqref{sec:Blind_And_Verified} by a toy version of a blind and verified computation protocols. Before concluding, the computational limitations of the toy model are examined in \eqref{sec:Computational_Limitations} along with some remarks about the implications of the existence of above mentioned toy stabilizer protocol. Finally, section \eqref{sec:Conclusion} summarizes the results and provides further lines of work.

\section{Background}
\label{sec:Background}

We start by reviewing the toy model  \cite{Spekkens2007}. States, transformations, and measurements in the toy model were defined by Spekkens in \cite{Spekkens2007}. Although Spekkens' notation provides a very natural and effective way to study properties of individual and pair of toy systems, it is inefficient at describing larger collections as such notation scales exponentially in the number of systems. In \cite{Pusey2012} Pusey introduced a new notation, reminiscent of the qubit stabilizer formalism \cite{Gottesman1997}, for the toy model. This notation scales polynomially in the number of systems and allows collections of more than three elements to be more easily treated. We will now firstly review Spekkens' and Pusey’s notations as the toy model is presented. We will then add several extensions including the partial trace operation, toy purifications, and irreversible transformations. These extensions will play an important role in our toy protocols and respective proofs. 

Much of what follows is built in direct analogy to the qubit stabilizer formalism, which we will now very briefly review (see~\cite{Gottesman1997} and Section 10.5 in~\cite{NielsenMichaelA.andChuang2010} for more details).
The qubit stabilizer formalism is built up using the Pauli group $P_1$, which is composed of the four Pauli operators $I$, $X$, $Z$ and $Y$ with all combination of phases $\pm 1, \pm i$. The n-fold Pauli group $P_n$  is given by all combination of tensor products of Paulis and again all combination of phases $\pm 1, \pm i$.
A qubit \emph{stabilizer group} $S$ is an abelian subgroup of the n-fold Pauli group which does not contain $-I{^\otimes n}$.
It can be efficiently described in terms of its $k$ generators $s_i \in S$ written as $S= \langle s_1,...s_k \rangle $, where $k\leq n$.
If there are $k=n$ generators, the stablizer group defines  the \emph{stablizer state} $ \ket{S} $ on $n$ qubits as the unique state satisfying the \emph{stabilizer equation}
\begin{eqnarray} 
\ket{S} = s \ket{S} ,\nonumber
\end{eqnarray}
for all $s \in S$. If there are fewer generators $k<n$, the set of states satisfying the stabilizer equation defines a subspace of dimension $2^k$. 
The state associated to the maximally mixed projection onto this space is given by
\begin{eqnarray}
\rho_S = \frac{1}{2^n}\sum_{s\in S} s.
\end{eqnarray}

\subsection{Toy States}
\label{sec:states}

The simplest allowed system in the toy model is always in one of four distinct possible \textit{ontic states}, represented by the vectors $\vec e_1 = (1,0,0,0)$, $\vec e_2 = (0,1,0,0)$, $\vec e_3 = (0,0,1,0)$, and $\vec e_4 = (0,0,0,1)$. The ontic states can be thought of as \textit{local hidden variables} which completely specify the state of the system but, crucially, are never directly accessible, preparable, or measurable by an observer. Instead what is accessible are the \textit{epistemic states}, which are a statistical description of the observer's state of knowledge of the ontic state. For example, an observer who knows a single toy system is either in the ontic state $e_{1}$ \textit{or} $e_{3}$ will assign epistemic state $e_{1} \vee e_{3}$ (i.e. read $e_{1}$ \textit{or} $e_{3}$) to such system. 

Crucial to the toy model is that, after imposing a restriction that limits the amount of information an epistemic state can carry about the underlying ontic variables, it is epistemic states, not ontic ones, which resemble quantum states. This constraint is formally introduced as the founding principle of the toy theory by Spekkens in~\cite{Spekkens2007} in the form of an epistemic restriction limiting allowed epistemic states. Toy epistemic states are always states of \textit{incomplete knowledge} and have to respect the so called `knowledge balance principle' (KBP). This principle quoted from \cite{Spekkens2007} reads ``\textit{If one has maximal knowledge, then for every system, at every time, the amount of knowledge one possesses about the ontic state of the system at that time must equal the amount of knowledge one lacks}". While to give a general example of the principle is somewhat akward (see~\cite{Coecke2011}), for a single system the KBP boils down to limiting the best allowed description of the system to one of the two ontic states, with equal probability. It is also important to note the reductionist approach of the toy model: the KBP limits more and more the allowed epistemic states as larger composite systems are considered.

Since the `$\vee$' notation introduced by Spekkens scales exponentially in the number of systems we will use the stabilizer like notation introduced by Pusey in~\cite{Pusey2012}. This is done by defining the following $4 \times 4$ diagonal matrices 

\begin{align}
\label{eq:ToyStabilizers}
\begin{split}
\mathcal{X} &= diag(1,-1,1,-1),\\
\mathcal{Y} &= diag(1,-1,-1,1),\\
\mathcal{Z} &= diag(1,1,-1,-1),\\
\mathcal{I} &= diag(1\hspace{0.15cm},1\hspace{0.15cm},1\hspace{0.15cm},1\hspace{0.15cm}).
\end{split}
\end{align}
These matrices are useful as they define a group, $G_1$, which plays an analogous role in the toy theory to the Pauli group  does in quantum theory. 
Note that throughout this paper calligraphy script such as $\mathcal{X}$ will be used to identify elements of the toy group $G$, whereas capital Latin script such as $X$ will be used for elements of the  Pauli group $P$ for the quantum case.
  
The matrices $\mathcal{X},\mathcal{Y},\mathcal{Z}$ represent the possible measurements on an elementary system. Considering a single toy system with ontic state $\vec e_{l}$, $l \in \{1,2,3,4\}$, a measurement of $g \in G_{1}$ will output a binary outcome $a = \{+1,-1 \}$ according to the eigenequation

\begin{equation}
\label{eq:ToyStabilizerPusey}
g \vec{e}_l = a \vec{e}_l.
\end{equation}

To go to $n$ systems, as in quantum mechanics, we take the tensor product. 
On the level of the ontic state we take tensor products of the vectors, for example a state of two systems both in state $\vec e_1$  is written $\vec{e_1}\otimes\vec{e_1}$. To represent epistemic states we use the tensor product  of elements of $G_1$.

The toy group $G_n$, where $n$ is the number of systems, is made up of all the $4^n$ by $4^n$ diagonal matrices of the form $\alpha g_1 \otimes \dots \otimes g_n$, with $\alpha = \{ 1, -1 \}$ and each $g_i \in G_{1}$. 
An element $g \in G_{n}$ is called a \textit{toy observable} or \textit{toy stabilizer}.
These operators represent measurements of $n$ toy systems and the measurement of $ g \in G_n$ also yields a binary output $a = \{+1,-1 \}$ according to the corresponding eigenequation 
\begin{equation}
g\vec{v} = a \vec{v},
\end{equation}
where $\vec{v} = \vec{e}_{l_1} \otimes \dots \otimes \vec{e}_{l_n}$ is an ontic vector of length $4^n$ specifying the state of $n$ elementary toy systems.

Epistemic states can now be introduced by using toy stabilizers.
Let the subgroup $S \subset G_n$ be such that $-\mathcal{I}^{\otimes n} \not\in  S$ and with all the elements of $S$ being mutually compatible, then $S$ represents a \textit{toy stabilizer group}. 
The \textit{toy stabilizer equations} for toy stabilizer group $S$ is the set of eigen-equations for underlying ontic state vectors $\vec{v}$
\begin{equation}
\label{eq:ToyEigenObservables}
s\vec{v} = \vec{v} \quad \text{for all} \,\; s \in S.
\end{equation} 
The epistemic state associated to a stabilizer group $S$ is defined as the one where any ontic state satisfying all the stabiliser equations is equally likely.

As in the quantum case, a toy stabilizer group $S$ can be most succinctly written in terms of a set of linearly independent elements, i.e. $S = \langle g_1, \dots , g_n \rangle$, where $H_S = \{ g_1, \dots , g_n \} $ is the generating set. 

In~\cite{Pusey2012} Pusey showed that these stabilizers defined epistemic states are complete in that all allowed epistemic states in the toy model can be written in this way by finding the suitable stabilizer group, and all stabilizer states are valid toy epistemic states (in particular they satisfy the knowledge balance principle), This explains also why it is not fundamental to formally introduce the KBP as one could \textit{a posteriori} simply use toy stabilizers as a way to define valid toy states.

The epistemic states are the main objects of interest in the toy model, since they are what can be accessed (prepared, measured), in particular for the running of our protocols. For this reason throughout the rest of the paper \emph{toy state} can be taken to mean \emph{toy epistemic state} - where confusion is possible the appropriate adjective will be added.

As it will be useful later, stabilizers can also be used to build diagonal matrices whose entries correspond to the probability of each individual ontic state~\cite{Pusey2012}. Given a toy stabilizer group $S = \langle g_1, \dots , g_n \rangle$, its ontic distribution corresponds to the diagonal $4^n \times 4^n$ matrix 

\begin{equation}
\label{eq:Ontic_Distribution}
\rho_{S} = \frac{\mid S \mid}{4^n} P_S =  \frac{1}{4^{n}}\prod_{g \in H_S} (\mathcal{I} + g),
\end{equation}
where $P_{S}$ is the projector onto the ontic states compatible with $S$ and is given by $P_{S} = (1/|S|) \sum_{s \in S} s $~\cite{Pusey2012}. In order to obtain~\eqref{eq:Ontic_Distribution} the projector has been rewritten with respect to the generating set by recalling that any element in the group can be written in terms of a product of generators, $s =  g_{1}^{a_{1}} \cdots g_{l}^{a_{n}}$ with $a_{i} = 0,1$. Indeed, $P_{S} = (1/|S|) \sum_{s \in S} s = (1/2^{n}) \sum_{a_{1}\dots a_{n}} g_{1}^{a_{1}} \cdots g_{l}^{a_{n}} = (1/2^{n})\prod_{j}^{n}(\mathcal{I} + g_{n})$. Furthermore, as the probability distribution needs to be normalized, $Tr(\rho_{S}) = 1$. The diagonal elements of $\rho_{S}$ represents the probability distribution of the underlying ontic states. 

The concepts of \emph{pure} and \emph{mixed} epistemic states can now be formally introduced. A toy epistemic state over $n$ systems is pure when its stabilizer group is composed of $n$ independent generators. Its matrix representation will then have $rank(\rho_{S}) = 4^{n/2}=2^{n}$, where the $2^{n}$ non zero diagonal entries take value $1/2^{n}$. By decreasing the number of independent generators mixed states can be obtained. If there are $l$ generators the rank is $rank(\rho_{S}) = 4^{n - l/2}$, with each entry equal. Its highest value is when $S = \mathcal{I}$, then the completely mixed state over $n$ systems is reached and $rank(\rho_{\mathcal{I}}) = 4^{n}$ while all entries of $\rho_{\mathcal{I}}$ equal $1/ 4^{n} $. Pure states can further be distinguished between `product states', and `maximally correlated'~\cite{Spekkens2007}. The maximally correlated ones represent `\textit{toy entangled}' states. E.g. $S = \langle \mathcal{X}\mathcal{X},\mathcal{Y}\mathcal{Y} \rangle$ is a `toy Bell state'.

The compatibility and incompatibility of the Pauli measurements in quantum mechanics is reflected by the statistics of measurements of the toy observables on the epistemic state. For example, for a single system the epistemic state associated to the toy statbiliser group generated by $\mathcal{X}$ will always return $+1$ if $X$ is measured, but if $\mathcal{Z}$ is measured it will equally likely return $+1$ or $-1$. Furthermore, the state after measurement is also disturbed. This is described in section \ref{sec:Measurements}. It will be helpful to define the following map which reflects
the analogy to the quantum case~\cite{Pusey2012}. Let $m: G_{n} \rightarrow P_{n}$ be the map suggested by the notation of equations~\eqref{eq:ToyStabilizers}, e.g. $m(\mathcal{Y} \otimes \mathcal{X}) = Y \otimes X$, then two toy observables $g$ and $h$ are compatible (incompatible) if $m(g)$ and $m(h)$ commute (anticommute).

Having seen the similarity between the toy and quantum cases, it is further important to explicitly give an example of the difference between quantum and toy stabilizers. Both Spekkens \cite{Spekkens2007} and Pusey \cite{Pusey2012} stress how the toy model here described is \textit{not} simply a restricted subtheory of quantum mechanics, but rather an entirely distinct model which reproduces many phenomenological aspects of quantum theory. In particular it is easy to see that there are quantum stabilizer state where there exists no one-to-one map to a single toy stabilizer. Take, for example, the quantum state stabilized by $S^{Q} = \{ XX,ZZ, -YY,II \}$. (Note, where there is a possibility for confusion we use superscript $Q$ for the quantum case and superscript $T$ for the toy case). This set does not define a valid toy stabilizer state as according to the definition of toy stabilizer~\eqref{eq:ToyStabilizers} we have that $\mathcal{X}\mathcal{Z} = \mathcal{Y}$ and hence $(\mathcal{X}\mathcal{X})(\mathcal{Z}\mathcal{Z}) = \mathcal{Y}\mathcal{Y}$. 

Rather than trying to match the full stabilizer sets quantum (toy) states can be related to toy (quantum) states by considering the different choices of generating sets. For example above the three equivalent generating sets for the $S^{Q}$ are $H_{1}^{Q} = \{ XX, ZZ\}$, $H_{2}^{Q} = \{ XX, -YY\}$, and $H_{3}^{Q} = \{ ZZ, -YY\}$. To each of these generating set an unique toy generator can be associated, and with that a unique toy epistemic state

\begin{align*}
H_{1}^{Q} \rightarrow H_{1}^{T} = \{ \mathcal{X}\mathcal{X}, \mathcal{Z}\mathcal{Z} \} &\Rightarrow S^{T}_{1} = \{ \mathcal{X}\mathcal{X}, \mathcal{Z}\mathcal{Z}, \mathcal{Y}\mathcal{Y}, \mathcal{II}  \}, \\
H_{2}^{Q} \rightarrow H_{2}^{T} = \{ \mathcal{X}\mathcal{X}, -\mathcal{Y}\mathcal{Y} \} &\Rightarrow S^{T}_{2} = \{ \mathcal{X}\mathcal{X}, -\mathcal{Z}\mathcal{Z}, -\mathcal{Y}\mathcal{Y}, \mathcal{II} \}, \\
H_{3}^{Q} \rightarrow H_{3}^{T} = \{ \mathcal{Z}\mathcal{Z}, -\mathcal{Y}\mathcal{Y} \} &\Rightarrow S^{T}_{3} = \{ -\mathcal{X}\mathcal{X}, \mathcal{Z}\mathcal{Z}, -\mathcal{Y}\mathcal{Y}, \mathcal{II} \}.
\end{align*}
The toy states identified by $S^{T}_{1},S^{T}_{2}$ and $S^{T}_{3}$ are not only all different but even mutually orthogonal as they contain stabilisers with opposite signs. Note, however, that the map between generators across the two theories is unique as the liner independence between the elements of a given generating set removes the sign ambiguities. 
We therefore relate a given quantum state $S^{Q}$ to a \textit{family} of toy states $\{S^{T}_{i}\}$, where the $i$ labels the a choice of generating for $S^{Q}$, and viceversa. 

We conclude by remarking that formally the difference arises from the fact that `toy-ly' $\mathcal{X}\mathcal{Z} = \mathcal{Y}$ while quantumly $XZ = -iY$, where again recall calligraphy script identifies toy stabilizers while plain script is reserved for quantum Pauli operators. This difference between the groups is also what in practice makes violations of Bell inequalities and any form of contextuality impossible in the toy model~\cite{Pusey2012}.

\subsection{Toy Reversible Transformations}
\label{sec:Transformations}

In the toy model both reversible and irreversible transformations can be defined. Here we review the former, firstly introduced in~\cite{Spekkens2007}, while we define the latter in section~\eqref{sec:Generalized_Transformations}.

Reversible transformations are represented by a $4^n$ by $4^n$ permutation matrix $\tilde{\mathcal{U}} \in U_{n}$ acting on the $4^n$ dimensional ontic vector $\vec{v}$, where $U_{n}$ is the permutation group over $n$ systems~\cite{Pusey2012}. 
In agreement with the KBP, permutations must be defined to take valid epistemic states to other valid epistemic states or, from a stabilizer perspective, toy stabilizer states to toy stabilzier states~\cite{Pusey2012}. 
A toy transformation $\tilde{\mathcal{U}}$ acting on the diagonal matrix $\rho_{S}$ produces a new state
\begin{equation}
\rho_{S' }= \tilde{\mathcal{U}} \rho_S \tilde{\mathcal{U}}^T.
\end{equation}

From a stabilizer perspective this can be seen as equivalent to updating the elements of the stabilizer group $S$ to a new stabilizer group $S^{\prime}$, which correctly describes the transformed state through

\begin{equation}
S^{\prime} = \tilde{\mathcal{U}} S \tilde{\mathcal{U}}^{T} = \langle \tilde{\mathcal{U}} g_1 \tilde{\mathcal{U}}^T, \dots , \tilde{\mathcal{U}} g_n \tilde{\mathcal{U}}^T \rangle. 
\end{equation}

For an elementary system there are $4!=24$ distinct allowed single system toy permutations~\cite{Pusey2012}. 
Operations between multiple systems are also allowed in so far as they take toy stabilisers states to toy stabiliser states and preserve the compatibility structure, see~\cite{Pusey2012} for more details on the allowed reversible transformations.

Toy permutations can be seen as analogous to the Clifford group in quantum theory~\footnote{The Clifford group is the group of unitary operations that maps the Pauli group to itself under conjugation.}. However, as with  states there is in general no unambiguous map between toy permutations and quantum Clifford unitaries. In particular, using the cycle notation~\footnote{e.g. the notation $(13)(24)$ means that $1 \rightarrow 3$, $3 \rightarrow 1 $ and $2 \rightarrow 4$, $4 \rightarrow 2$.} we can identify two classes of permutations:

\begin{itemize}
	\item The permutation subgroup identified by the single system two-cycles $\tilde{\mathcal{X}} = (12)(34)$, $\tilde{\mathcal{Y}} = (14)(23)$, $\tilde{\mathcal{Z}} = (13)(24)$ and the identity $()$ as it has an unambiguous map to the Pauli subgroup of quantum Clifford operations. That is, they are in a one-to-one correspondence with the quantum Pauli operators $Z,Y,X$ and $I$ respectively. We call this special group of permutations the toy Pauli permutation subgroup,
	\item An arbitrary toy permutation $\tilde{\mathcal{U}}_{i} \in U_{n}$ which is not composed a tensor product of single system toy Pauli operations cannot unambiguously be mapped to a Clifford operation, and viceversa.
\end{itemize}

To understand why there is no one-to-one map between arbitrary permutations in the two theories note how the difference between the toy $\mathcal{X}\mathcal{Z} = \mathcal{Y}$ and quantum $XZ = -iY$ implies that transformations will not in general have the same effect on all elements of the respective stabilizers. For example a permutation with a consistent action on both toy $\mathcal{X}, \mathcal{Z}$ and quantum $X, Z$ states will not implement the same transformation in both theories on the third Pauli `$Y$' state.

Let us now suggest a systematic way to map operations between the two theories. As it is not possible to define an unique map, we will define a standard map between quantum and toy operations with respect to the action of the permutation/unitary over the pairs $(\mathcal{X},\mathcal{Z})$ and $(X,Z)$. Explicitly let $f: U_{n} \rightarrow C_{n}$, where $C_{n}$ is the Clifford group over $n$ systems. Then a toy operation $\tilde{\mathcal{U}} \in U_{n}$ is mapped to a quantum Clifford operation $C = f(\tilde{\mathcal{U}}) $ by imposing the condition that $m( \tilde{\mathcal{U}} \mathcal{A} \tilde{\mathcal{U}}^{T} ) = C A C^{\dagger}$ where $m$ is the map introduced in section~\eqref{sec:states} for all $\mathcal{A}$ ($A$) that are tensor products of $\mathcal{I}$s, $\mathcal{X}$s, and $\mathcal{Z}$s ($I$s, $X$s, and $Z$s). 
The map $f$ defined with respect to $\mathcal{X}$ and $\mathcal{Z}$ is taken as standard and it will be used to relate operations between quantum and toy protocols. Note however such map could have been defined with respect to any another pair $(g_{i},g_{j})$ of single system toy stabilizers $g_{i},g_{j} \in G_{1} \setminus \mathcal{I}$ and $g_{i} \not = g_{j}$. 

To conclude we remark on the existence of entangling operations in the toy model. In order to respect the KBP principle only controlled Pauli permutations are allowed. Indeed this is analogous to Clifford in the quantum case. We will explicitly define and use the control-$\tilde{\mathcal{Z}}$ in order to create the graph states needed for the protocols in section~\eqref{sec:Results}.

\subsection{Toy Measurements}
\label{sec:Measurements}

Measurements in the toy model are probabilistic processes that partition the underlying ontic space into valid epistemic states \cite{Pusey2012}. They are also invasive operations in the sense that by the act of performing a measurement the ontic state of the measured system is disturbed, and its corresponding new epistemic state needs to change consistently~\cite{Spekkens2007}. This is again a consequence of the KBP or, from the stabiliser perspective, reflects the need of obtaining a valid stabiliser state after the measurement has been performed. 

The most general way to write a measurement over $n$ toy systems is  

\begin{equation}
\label{eq:GeneralizedMeasurements}
M = \sum_{i} \alpha_i P_{S_{i}},
\end{equation}
where each $P_{S_{i}} = 1/|S_{i}| \sum_{s_{j} \in S_{i}} s_{j}$ corresponds to a projector onto the state or space associated to stabilizer group $S_{i}$, while each $\alpha_{i}$ labels the outcome associated to the $i$-th partition of the measurement. 
In order for $M$ to be a valid measurement the union of the partitions $S_{i}$ must correspond to a partition of unity. That is, they must be mutually disjoint and cover the whole of the ontic space:
\begin{equation}
\label{eq:Measurements_Validity_Condition}
\sum_{i} P_{S_{i}} = \mathcal{I}^{\otimes n}.
\end{equation}

Any set $\{ S_{1}, \dots \}$ of stabilizers satisfying the above condition thus defines a valid measurement through~\ref{eq:Measurements_Validity_Condition}.

Given a reference system $\rho_{R}$ the probability of getting outcome $\alpha_{i}$ is given by 
\begin{equation}
\label{eq:Measurement_Outcomes}
prob(\alpha_{i}) = Tr (P_{S_{i}} \rho_{R}).
\end{equation}

Fundamentally, this equation tells us the probability that the ontic state of the reference system $\rho_{R}$ is compatible with the $i$-th partition of the measurement $M$.

The resulting state is found as follows. Consider performing a measurement $M = \sum_{i} \alpha_i P_{S_{i}}$ on the state $\rho_{R}$ with stabilizer group $R = \{ r_{1}, \dots, r_{|R|} \}$. Each outcome $\alpha_{i}$ is then possible with probability $Tr(P_{S_{i}} \rho_{R})$. Assuming we observe the $\alpha_{i}$ outcome, then consistency implies the stabilizer of the post-measured state must be updated so to contain the stabilizer group $S_{i}$ plus all the stabilizers of $R$ which are compatible with the measured $S_{i}$. Explicitly, the full stabilizer of the new state is

\begin{equation}
\label{eq:Post_Measurement_Update_Rule}
R^{\prime} = \{ \{ S_{i}\} , \{ r_{j} \; | \; r_{j} \in R \text{ and } [m(r_{j}),m(s_{m})] = 0 \; \forall \; s_{m} \in S_{i} \} \}, 
\end{equation}
where $m$ is the map to quantum operators defined in section \eqref{sec:states}, indicating that $r_j$ and $s_m$ are compatible.

In the case where the measurement itself is a toy operator $g \in G_{n}$ (i.e. $M=P_{+g}-P_{-g}$), the update can be simply described with Paulis as in the quantum case.
Consider therefore measuring the toy observable $g \in G_{n}$ on a state stabilized by $R$. If $\pm g \in R$, then no change on the resulting epistemic state is effected and $R^{\prime} = R$; otherwise, the resulting stabilizer $R^{\prime}$ will be build from the measured $\pm g$ and all the elements of $R$ which commute with it according to the map defined in~\eqref{sec:states}. In practice this is done by listing all the generators of $R$ which are incompatible with $\pm g$. If only one is incompatible, then $\pm g$ replaces such generator hence obtaining the new toy stabilizer state $R^{\prime}$. If more than one generator is incompatible, then by multiplying the incompatible generators one with another a list of independent generators for the old state with at most one element that anticommutes with $\pm g$ is obtained. Then just replace it by $\pm g$ to obtain the new state~\cite{Pusey2012}.

If the total measurement does not correspond to a single toy observable $g \in G_{n}$ but it is rather described as a toy stabilizer $S \subset G_{n}$, one has to treat different projections differently. In each case one can simply choose any single toy observable which contains the projection (for example the stabiliser of the projection itself) and proceed as above.

We further stress that the only requirement for a valid measurement is given by~\eqref{eq:Measurements_Validity_Condition}. This is in line with the observation in~\cite{Pusey2012} as not all partitions satisfying equation~\eqref{eq:Measurements_Validity_Condition} necessarily correspond to toy observables. Finally, we note a new, more formal, update rule has been recently presented in~\cite{Catani2017}
 
\subsection{Toy purifications}
\label{sec:Toy_Purifications}

As mentioned in the introduction, when considering composite toy systems it is important to note the `reductionist' flavor of the toy theory: since any part of the collection must individually correspond to a valid toy state, the KBP must hold at any scale of the composite system \cite{Spekkens2007}. This does not only imply the existence of a well defined partial trace operation, but also provides a meaningful way to introduce the concept of purifications in the toy model. Informally, the ability to purify a state corresponds to the ability to describe any mixed state over some collection of systems $A$ as the marginal state arising from a pure state defined in some larger collection of systems, composed by $A$ and some extra set of reference systems $R$. 

We here prove the existence of purifications in the toy model and their equivalence up to permutations on the reference systems $R$, hence mimicking two known results of quantum theory \cite{NielsenMichaelA.andChuang2010}. 

To prove our results we will build an explicit map between quantum and toy states such that the existence of stabilizer purifications in quantum theory implies the existence of toy ones. Without loss of generality consider a bipartite toy system $\rho_{AR}$ stabilized by $S_{AR}$ where subscripts $A$ and $R$ label two disjoint set of physical systems, each composed by $n_{a} = n_{r}$ physical systems. The partial trace is then defined with respect to either $A$ or $R$ and is a procedure which yields the marginal ontic distribution of either set of systems with respect to their joint distribution. In the toy model the partial trace cannot be simply obtained from the generators but is a property of the full stabilizer group $S$:
\begin{align}
\label{eq:Trace_out_Stabilzier_Defn}
Tr_{R}(S_{AR}) &= Tr_{R} (\{ s_{1}, \dots, s_{|S_{AR}|} \}) \nonumber \\
&= \{ a_{A} \; | \; a_{A} \otimes \mathcal{I}_{R} \in S_{AR} \} \nonumber \\
&= S_{A}.
\end{align}

Fundamentally, it is important to observe that the only contributions to the partial trace are from those stabilizers in $S_{AR}$ of the form $ \pm a_{A} \otimes \mathcal{I}_{R}$, as the identity is the only toy stabilizer with non trivial trace i.e. $tr{(\mathcal{I})} = 4$.

In terms of the ontic distributions, the partial trace can, analogously to the quantum case \cite{NielsenMichaelA.andChuang2010}, be represented by

\begin{equation}
\label{eq:Trace_Out_Definition}
Tr_{R}(\rho_{AR}) = \frac{1}{4^{n}} Tr_{R}(\sum_{s \in S_{AR}} s) = \rho_{A}.
\end{equation}

We are now ready to make the following purification statement.

\begin{theorem}
\label{thm:Existence_of_Purifications}
\textit{To any mixed epistemic state $\rho_{A}$ defined over a set of systems $A$ there exists a pure epistemic state $\rho_{AR}$ defined over $A$ and an auxiliary set of reference systems $R$ of at least equal size, such that the marginals of $A$ in $\rho_{AR}$ are original state $\rho_{A}$.}
\end{theorem}

\textit{Proof.} 
Using superscripts $T$ and $Q$ to label toy and quantum states respectively, we want to show that for an arbitrary mixed toy state $\rho^{T}_{A}$ we can find a mixed quantum state  $\chi^{Q}_{A} \leftarrow \rho^{T}_{A} $ with its respective quantum stabilizer purification $\chi^{Q}_{AR}$, i.e. where $Tr_{R}(\chi^{Q}_{AR}) = \chi^{Q}_{A}$, and from the quantum purification we can further identify a pure toy state $\rho^{T}_{AR} \leftarrow \chi^{Q}_{AR} $ over $AR$ such that it is a valid toy purification of the original mixed toy state, i.e. $Tr_{R} (\rho^{T}_{AR}) = \rho^{T}_{A}$. 

The difficulty here is not only given by the lack of a unique map between toy and quantum states, but also because the chosen map must induce the correct `purification-relationship' between the toy states $\rho^{T}_{AR} $ and $ \rho^{T}_{A}$.

Let us then suggest the following rule to ensure consistency. Since the only terms surviving the partial trace in equation~ \eqref{eq:Trace_out_Stabilzier_Defn} correspond to a subset. In a slight abuse of notation we write $S_{A}^{Q} \subseteq S_{AR}^{Q}$ (in reality $S_A$ is defined on fewer systems so it is not strictly a subset, but its non-trivial parts stem from subset of $S_{AR}^{Q}$ via equation~ \eqref{eq:Trace_out_Stabilzier_Defn}. Then there there exists a choice of generating set $G_{A}^{Q}$ for $S_{A}^{Q}$, which can be used to define an explicit generating set of $S_{AR}^{Q}= \langle G_{A}^{Q}, \dots  \rangle$ for the joint system, where the other generators are arbitrarily chosen to complete the set. That is, because the partial trace simply removes all the stabilizers which are non trivial over the reference system, a valid purification $S_{AR}^{Q}$ must have an explicit choice of generating set such that it includes $G_{A}^{Q}$.

To any mixed toy epistemic state $\rho_{A}^{T}$ described by stabiliser group $S_{A}^{T} = \langle G_{A}^{T} \rangle$ we can associate a pair of quantum states: a corresponding quantum stabiliser state $S_{A}^{Q} = \langle G_{A}^{Q} \rangle \leftarrow S_{A}^{T}$ and its respective quantum purification $S_{AR}^{Q} =< \{ G_{A}^{Q} \}, \dots > $. When choosing the generating set for the quantum purification $S^{Q}_{AR}$ consider the one(s) explicitly featuring the generators of the state stabilized by $S_{A}^{Q} = \langle G_{A}^{Q} \rangle$. (Since $S_{A}^{Q}$ is contained in $S_{AR}^{Q}$ this is always possible). We can then identify a candidate toy purification by defining the toy stabilizer group $S_{AR}^{T} = \langle \{ G_{A}^{T} \}, \dots \rangle \leftarrow S_{AR}^{Q} = \langle \{ G_{A}^{Q} \}, \dots \rangle$. Now, when the partial trace is performed on the candidate toy purification the only terms in $S^{AR}_{T}$ contributing are those contained in $G_{A}^{T}$ which is the generating set of the original toy state $\rho_{A}^{T}$. This is because both in the toy and quantum case the only stabiliser with non vanishing trace is the identity. \qed

\begin{figure}
\begin{equation*}
\begin{tikzpicture}
\draw (-2,0) node[minimum height=1cm,minimum width=1cm] (A) {${S_{A}^{T} = \langle G^{T}_{A} \rangle }$};
\draw (4,0) node[minimum height=1cm,minimum width=1cm] (B) {$S_{AR}^{T} = \langle \{ G_{A}^{T} \}, \dots \rangle $};
\draw (-2,2) node[minimum height=1cm,minimum width=1cm] (C) {${S_{A}^{Q} = \langle G^{Q}_{A} \rangle }$};
\draw (4,2) node[minimum height=1cm,minimum width=1cm] (D) {$S_{AR}^{Q} = \langle \{ G_{A}^{Q} \}, \dots \rangle $};

\draw[<->,red] (A) -- node[above] {$Tr_{R}$} (B);
\draw[->] (A) -- (C);
\draw[<->] (C) --  node[above] {$Tr_{R}$}(D);
\draw[->] (D) -- (B);
\end{tikzpicture}
\end{equation*}
\caption{\footnotesize A schematic representation of theorem~\eqref{thm:Existence_of_Purifications}. Start from the bottom left with the mixed toy state stabilized by $S_{A}^{T} = \langle G^{T}_{A} \rangle$. This identifies a pair of quantum states, the mixed one stabilized by $S_{A}^{Q} = \langle G^{Q}_{A} \rangle$ and its purification who's generator is explicitly written as $S_{AR}^{Q} = \langle \{ G_{A}^{Q} \}, \dots \rangle $. Note the generating set of the quantum purification has been chosen in order to explicitly feature the generators $G_{A}^{Q}$. The toy purification identified as $S_{AR}^{T} = \langle \{ G_{A}^{T} \}, \dots \rangle   \leftarrow S_{AR}^{Q} = \langle \{ G_{A}^{Q} \}, \dots \rangle >$ is then also a valid purification of the original toy state.}
\label{fig:Purifications}
\end{figure}
\medskip

Graphically these choices can be viewed in figure~\eqref{fig:Purifications}. The key idea here is that the ambiguity between toy and quantum states can be focused on the part of the purified state which is not relevant to the properties needed to define the operation. Further, see that all toy purifications are equivalent up to a permutation acting only on the reference system $R$.

\begin{theorem}
\label{thm:Equivalence_Of_Purifications}
\textit{Let $\rho_{AR}$ and $\sigma_{AR}$ be two toy purifications of the same toy state $\rho_{A}$. Then there exists a toy permutation of the form $\tilde{\mathcal{I}}_{A} \otimes \tilde{\mathcal{U}}_{R}$ such that $(\tilde{\mathcal{I}}_{A} \otimes \tilde{\mathcal{U}}_{R}) \sigma_{AR} (\tilde{\mathcal{I}}_{A} \otimes \tilde{\mathcal{U}}_{R})^{T} = \rho_{AR} $.}
\end{theorem}

\textit{Proof.} 
As with the proof above, one can translate the quantum result over. Any two quantum states defined over two systems, A and B, with identical reduced states A can be related by unitaries on B. Moreover, if these states are stabiliser states, this unitary can be chosen to be Clifford. By choosing appropriate maps from the toy to quantum cases, as above, one sees that the toy version is related by local permutation on B. \qed

\subsection{Toy Generalized transformations}
\label{sec:Generalized_Transformations}

We here introduce a class of transformations not yet defined within the toy model. These are the irreversible transformations and are analogous to the quantum completely positive trace preserving maps (CPTP). We model these, inspired by the quantum case, by first entangling the system of interest with an ancilla system, then performing reversible transformations and measurements over the system plus ancilla, followed by a partial trace over the ancilla systems.

Let $\rho_{A}$ be the state of system $A$ and $\sigma_{R}$ the state of the ancilla. Their joint system is in a product state of the form $\rho_{A} \otimes \sigma_{R}$. 
Let then $\rho_{A} \otimes \sigma_{R}$ be evolved by means of a global permutation $\tilde{\mathcal{ U }}_{AR}$ giving rise to 

\begin{equation}
\sigma_{AR_S} = \tilde{\mathcal{U}}_{AR} (\rho_{A} \otimes \sigma_{R})  \tilde{\mathcal{U}}_{AR}^{T},
\end{equation}
where we subscript $S$ labels the respective stabilizer group. Next, a measurement on the auxiliary systems $R$ alone is performed. Such measurement can be described by $M = \sum_{i} q_{i} I_{A} \otimes P_{R_{T_{i}}}$ such that $\sum_{i} P_{R_{T_{i}}} = I_{R}$. 

We get the result $q_i$ with probability $prob(q_{i})$, resulting in state $\chi_{AR}^{S^{\prime}_{i}}$ (with stabiliser group $S^{\prime}_{i}$), We express this as the ensemble
\begin{equation}
\{ prob(q_{i}), \; \chi_{AR_{S^{\prime}_{i}}} \}.
\end{equation}

The generators of the stabilizer groups $S^{\prime}_{i}$ are obtained, as shown in section~\eqref{sec:Measurements}, starting from the generators identified by the $T_{i}$ partition, and then adding a set of compatible and linearly independent generators of the former $\sigma_{AR_{S}}$  state. However, as we are interested in the description of this state from the perspective of $A$ we further trace out all subsystem belonging to $R$. This generates a new ensemble of states on $A$ alone

\begin{equation}
  \{ prob(q_{i}), \; \chi_{A}^{S^{\prime \prime}_{i}} = Tr_{R}(
  \chi_{AR}^{S^{\prime}_{i}}) \}.
\end{equation}  
Knowledge of the specific observed $q_{i}$ then selects a single state from the ensemble. If no information about the result is provided, we can just average over the possible resulting states and is given by

\begin{equation}
\chi_{A} = \sum_{i} \alpha_{i} \chi_{A}^{S^{\prime\prime}_{i}},
\end{equation}
where $\alpha_{i} =  prob(q_{i})$. This is a valid state since it is essentially equivalent to simply tracing out $R$ on the unmeasured state. 

However, if one has partial information about the results, it is not clear if the analogous object would be a valid state, just as in the quantum case such state would not need to be a stabilizer state. Indeed arbitrary mixtures of stabilizer states are \textit{not} allowed in the toy model~\cite{Spekkens2007,Pusey2012}. Thus one has to treat the role of this classical information with some caution. 

This issue will also arise in the context of the protocols for blind and verified computation~\eqref{sec:Blind_And_Verified}, where Bob is ignorant of some classical information. If one described this situation as a statistical mixture over the unknown classical information, from Bob's perspective the state may appear to be disallowed in the toy theory
(though from Alice's point of view, with the information, everything is valid). In this sense these operations may lead to extensions of the toy theory. We will delay discussion on how this addition changes the toy model in section~\eqref{sec:Computational_Limitations} where we will consider the explicit case of blind and verified computations.

\section{Toy protocols}
\label{sec:Results}

\subsection{Recovering the commutation relations}
\label{sec:Recovering_Commutation_Relations}

In quantum mechanics Pauli operators play a double role: they can be used to represent either a  measurement on a system  or a unitary transformation. However, measurements and transformations in the toy theory are represented by two distinct mathematical objects which individually do not share the standard quantum commutation relations, in particular $\tilde{\mathcal{Z}} \tilde{\mathcal{X}} = \tilde{\mathcal{X}} \tilde{\mathcal{Z}}$ and $\mathcal{X}\mathcal{Z} = \mathcal{Z}\mathcal{X}$. The first step in order to define more complex toy protocols than those defined in \cite{Spekkens2007} is to operationally recover the correct (anti)commutation relations. To this end, we observe that toy observables $\mathcal{X}$ ($\mathcal{Z}$) anticommute with the toy permutations $\tilde{\mathcal{Z}}$ ($\tilde{\mathcal{X}}$). That is, $\tilde{\mathcal{Z}} \mathcal{X} = - \mathcal{X} \tilde{\mathcal{Z}}$ and $\mathcal Z \tilde{\mathcal{X}} = - \tilde{\mathcal{X}} \mathcal Z$. This almost trivial observation will play a key role in making the protocols in the next sections work.

These relations also help to define two new sets of stabilizers equations. One in terms of toy permutations and the other in terms of toy measurements. These two new sets of equations are equivalent to the toy stabilizer equations~\eqref{eq:ToyEigenObservables}, but can be more effectively used to define quantum protocols in the toy model.

To obtain the set of toy-permutation stabilizer equations define a map $h: G_{n} \rightarrow U_{n}$ such that to any set of toy stabilizers associates a set of toy Pauli permutations, e.g. $h(\mathcal{Z} \otimes \mathcal{Y}) \rightarrow \tilde{\mathcal{Z}} \otimes \tilde{\mathcal{Y}}$. Then given a stabilizer $S$, $h$ identifies a permutation stabilizer $\tilde{S} = h(S)$ (where a function on a set is taken to mean the set produced by applying the function on each element). Note that this map is unambiguous, and all the permutations obtained in this way are part of the permutation subgroup previously introduced as the Pauli permutation subgroup. The action of $\tilde{S}$ leaves then the stabilized state $\rho_{S}$ unchanged:

\begin{equation}
\label{eq:Toy_Permutation_Stabilizers}
\tilde{s} \rho_{S} \tilde{s}^{T} =  \rho_{S}, \quad \forall \; \tilde{s} \in \tilde{S}.
\end{equation}

This can be easily verified by noticing that $\tilde{s}_i s_j = s_j \tilde{s}_i$ for all $s_{j} \in S$. 

In order to obtain the set of toy-measurement stabilizer equations note how equations~\eqref{eq:ToyEigenObservables} can be rewritten in terms of measurements as follows. 
$\forall g \in S$,
\begin{equation}
\label{eq:Stabilizer_Trace}
Tr(P_{g} \rho_{S}) = 1 \Longleftrightarrow  g \vec{v} = \vec{v}\;  \text{ (and }  \vec{v} \in supp(P_{g})).
\end{equation}

That is, the probability of obtaining the $+1$ outcome upon measuring the toy observable $g$ on the state $\rho_S = \frac{1}{4^n} \sum_{g \in S} g$ must equal unity as the measured observable belongs to the state's stabilizer group. Therefore equations~\eqref{eq:Toy_Permutation_Stabilizers} and~\eqref{eq:Stabilizer_Trace} can independently be used to describe a state, and provide an alternative formulation of the stabilizer equations in~\cite{Pusey2012}.


\subsection{Encoding information}
\label{sec:Encoding_Information}

A first use for the new stabilizer equations \eqref{eq:Toy_Permutation_Stabilizers} and \eqref{eq:Stabilizer_Trace} is to show how $k$  systems can be encoded within $n > k$ systems. The procedure is parallel to the one used to encode states into a quantum stabilizer state~\cite{Gottesman1997,NielsenMichaelA.andChuang2010}.

Our goal is to define a map from the `physical' $k$ systems to the `logical' $n$ systems  for states,  operations, and measurements in a consistent way. That is, for any state $\rho$,  reversible transformations $\tilde{\mathcal{U}}$ and measurement $M=\sum_i \alpha_i P_{S_i}$ on the $k$ systems, we want a map to $n>k$ systems $\rho \rightarrow \rho_L$,  $\tilde{\mathcal{U}} \rightarrow \tilde{\mathcal{U}}_L$ and $M \rightarrow M_L =\sum_i \alpha_i P_{S_i,L}$ (we use subscript $L$ to indicate the logical encoding), such that
\begin{align}
\label{eq:EncodingConditions}
Tr(P_{S_i} \rho_{S}) &= Tr (P_{S_i,L} \rho_{S_{L}} ), \\
\textit{If } \rho_{S}^{\prime} = \tilde{\mathcal{U}} \rho_{S} \tilde{\mathcal{U}}^T &\textit{, Then } \rho_{S_{L}}^{\prime} = \tilde{\mathcal{U}}_{L} \rho_{S_{L}} \tilde{\mathcal{U}}^T_{L}. 
\end{align}
 
This is done using stabilisers as follows. Over $n$ systems one defines the logical space using generators $g_1,..., g_{n-k}$.  Next, one defines a complete set of logical operators, that is for every Pauli $h$ on $k$ systems, the logical version $h_L$, which effectively chooses a basis in this space.
Then for the stabiliser group on k systems $S=\langle h_1, ...h_k\rangle$ defining a state, the logical encoding on $n$ systems is $S_L = \langle g_1, ...,  g_{n-k}, h_{1,L}, ... h_{k,L}\rangle$.

\subsection{No deletion and no bit-commitment }
\label{sec:Bit_Commitment_No_Delition_Theorem}

The existence of purifications and their equivalence up to a local permutation on the reference system bears great importance for many information processing tasks. We here prove a toy version of the no deletion theorem and the impossibility of (im)perfect toy bit commitment schemes, hence reproducting two known results of quantum theory, see~\cite{Pati2000} and~\cite{Lo1997,Mayers1997,Leverrier2014} respectively.

\subsubsection{No deletion theorem}
\label{sec:No_Deletion}
The idea of the no deletion theorem is that information cannot be deleted, rather only moved about. Informally, if information (a bit, a qubit, or a single toy system) is encoded over two systems, if one system has no access to the information, the other system has perfect access. 

Start by considering the encoding of one single toy system into $n$ physical toy systems. Let the $n$ systems be divided into two distinct regions, $A$ and $B$, each containing $n_{a}$ and $n_{b}$ physical systems respectively with $n = n_{a} + n_{b}$. The resulting encoded state is stabilized by $S_{L}  = \langle g_{1}, \dots, g_{n-1}, h_{L} \rangle $ where $h_{L} \in \{ \pm \mathcal{X}_{L}, \pm \mathcal{Z}_{L}, \pm \mathcal{Y}_{L} \}$ represents the logical state of the encoding. When the logical encoding is initialized in the `zero' logical state (i.e. $+\mathcal{Z}_{L}$) it will be labelled $\rho^{S_{0}}_{AB}$ and similarly for the other logical states i.e. states $\rho^{S_{1}}_{AB}$, $\rho^{S_{+}}_{AB}$, $\rho^{S_{-}}_{AB}$, $\rho^{S_{+i}}_{AB}$, $\rho^{S_{-i}}_{AB}$ correspond to generators , $-\mathcal{Z}_{L}$, $\mathcal{X}_{L}$, $-\mathcal{X}_{L}$, $\mathcal{Y}_{L}$ and $-\mathcal{Y}_{L}$ respectively.

Now consider the case where the reduced states over $B$ be are such that

\begin{align}
Tr_{A}(\rho^{S_{0}}_{AB}) &= \rho_{B} = Tr_{A}(\rho^{S_{1}}_{AB}), \label{eq:Equivalence_Logical_States_1} \\
Tr_{A}(\rho^{S^{+}}_{AB}) &= \sigma_{B} = Tr_{A}(\rho^{S^{-}}_{AB}). \label{eq:Equivalence_Logical_States_2}
\end{align}

That is, the two logical orthogonal states $\rho^{S_{0}}_{AB}$ ($\rho^{S_{+}}_{AB}$) and $\rho^{S_{1}}_{AB}$ ($\rho^{S_{-}}_{AB}$) correspond to the same reduced state when $A$ is traced out. This means that $B$ cannot tell whether his reduced state $\rho_{B}$ (or $\sigma_{B}$) is part of the `zero' (`plus') or `one' (`minus') logical state. Furthermore, by theorem~\eqref{thm:Equivalence_Of_Purifications} the logical states over $AB$ must be related one to another by a local permutation on $A$. As these local permutations relate logical states they are logical operations and labeling them $\tilde{\mathcal{X}}_{AB_{L}} = \tilde{\mathcal{X}}_{A} \otimes \tilde{\mathcal{I}}_{B}$ and $\tilde{\mathcal{Z}}_{AB_{L}} = \tilde{\mathcal{Z}}_{A} \otimes \tilde{\mathcal{I}}_{B}$ respectively, their action on the logical state can be explicitly written as

\begin{align}
\rho^{S_{0}}_{AB} &=  (\tilde{\mathcal{X}}_{A} \otimes \tilde{\mathcal{I}}_{B}) \rho^{S_{1}}_{AB} (\tilde{\mathcal{X}}_{A} \otimes \tilde{\mathcal{I}}_{B})^{T}, \label{eq:No_Delition_Theorem_1}\\
\rho^{S_{+}}_{AB} &= (\tilde{\mathcal{Z}}_{A} \otimes \tilde{\mathcal{I}}_{B}) \rho^{S_{-}}_{AB} (\tilde{\mathcal{Z}}_{A} \otimes \tilde{\mathcal{I}}_{B})^{T}. \label{eq:No_Delition_Theorem_2}
\end{align}

In this way $A$ has access to the encoded information - she can change it and read it as she likes.

Similar statements can also be made with respect to the ability of extracting the encoded information. That is, the probability associated to a measurement $M = \sum_{i} q_{i} I_{A} \otimes P_{B_{T_{i}}}$ on system $B$ alone must be completely independent from the encoded state, therefore for any $T_{i}$

\begin{align}
 Tr((I_{A} \otimes P_{B_{T_{i}}}) &\rho^{S_{j}}_{AB}) =  Tr((I_{A} \otimes P_{B_{T_{i}}}) \rho^{S_{k}}_{AB}) , \nonumber \\
  &\textit{ for } j ,k \in \{0,1,+,-\}.
\end{align} 

All the above equations are fundamentally statements about the localization of the encoded information. It was already proven by Spekkens in \cite{Spekkens2007} that the toy model features a no cloning theorem, we show here that the toy model further exhibit an `inverse no cloning', also known as the no-deletion \cite{Pati2000}. In essence equations~\eqref{eq:Equivalence_Logical_States_1} and ~\eqref{eq:Equivalence_Logical_States_2} imply that if the partition $B$ has no information about the logical encoding, then the partition $A$ has full access over the encoding. Full access over the encoding means that the logical encoded state is perfectly \textit{measurable} and \textit{manipulable} by toy stabilizers and permutations defined over $A$ alone.
In particular, in the case of a bipartite system this implies that the complement has all the information. This is the content of the no deletion theorem~\cite{Pati2000}. The result easily generalizes to an arbitrary pure encoding of $k$ logical toy systems within $n>k$ physical toy systems.

\subsubsection{No (im)perfect bit commitment}

Weakening the conditions~\eqref{eq:Equivalence_Logical_States_1} and~\eqref{eq:Equivalence_Logical_States_2} by demanding only one of them, say equation~\eqref{eq:Equivalence_Logical_States_1}, to hold corresponds to the standard scenario for proving the impossibility of perfect bit commitment in quantum theory~\cite{Lo1997,Mayers1997}. 

In its simplest form a bit commitment protocol involves two parties $A$(lice) and $B$(ob), a single committed bit of information $b$, and a single round of communication. More involved protocols with longer commitment strings and many concatenated rounds of communication can also be considered, but to prove the impossibility of the scheme it suffices to consider one single bit and one round of communication~\cite{Lo1997,Mayers1997}.

In this way, without loss of generality any bit commitment protocol can be divided into three stages:

\begin{enumerate}
	\item \textit{Commitment phase}: Alice is asked to commit to a single bit $b \in \{0,1\}$.
	\item \textit{Storage phase}: Alice encrypts the bit via a secret key. While she keeps the key secret, the encoded message  is sent to Bob who stores it for an arbitrary amount of time.
	\item \textit{Revealing phase}: After the commitment time has elapsed,  Alice sends the decoding key to Bob who can now apply the key to the message and check the value Alice had previously committed to.
\end{enumerate}

There are two requirements for a \textit{perfect} bit commitment scheme: firstly once the commitment phase is ended there is no way for Alice to alter her choice, and, secondly, during the storage phase Bob cannot obtain any information about the value of $b$. It is known that neither classically nor quantumly it is possible to satisfy both criteria at the same time~\cite{Lo1997,Mayers1997}.

Before proving the impossibility of bit commitment in the toy model, let us review how bit commitment works in the quantum case. From the perspective of making statements about security, one can phrase the most general protocol as follows~\cite{Lo1997,Mayers1997}. In the \textit{commitment} phase  Alice encodes one bit $b \in \{0,1\}$ into a (possibly entangled) state $\rho_{AB}^{b}$, and sends half of it to Bob. As the state reaches Bob the commitment phase is concluded and the \textit{storage} phase begins. For a perfect scheme, Bob's reduced state $\rho_{B}^{b}$ during the storage is such that $Tr_{A}(\rho^{0}_{AB}) = \rho_{B} = Tr_{A}( \rho^{1}_{AB})$. The state is held by Bob until the \textit{revealing} phase, when Alice sends her part of the state $\rho_{AB}^{b}$ thus enabling Bob to unveil the value of the commitment. 

Returning now to the Toy case, we now show that in the toy model is not possible to have a perfectly secure toy bit commitment scheme.

\begin{theorem}
\textit{Perfect toy bit commitment is impossible in the toy model.}
\label{thm:Perfect_Bit_Commitment}
\end{theorem}

\textit{Proof.} Let Alice encode one single system into $n$ physical systems and obtain $S_{b} = \langle g_{1}, \dots, g_{n-1}, h_{L_{b}} \rangle$. The committed state is represented by $\rho^{S_{b}}_{AB}$. The first requirement for perfect bit commitment is that during the storage part Bob has no information about the committed state. That is, $Tr_{A}(\rho^{S_{0}}_{AB}) = \rho_{B} = Tr_{A}(\rho^{S_{1}}_{AB})$. This implies the states $\rho^{S_{0}}_{AB}$ and $\rho^{S_{1}}_{AB}$ are both purifications of $\rho_{B}$, and by theorem~\eqref{thm:Equivalence_Of_Purifications} they are related one to another by a permutation of the form of  equation~\eqref{eq:No_Delition_Theorem_1}. Hence Alice can perfectly change her commitment during the storing stage by locally applying the logical permutation $\tilde{\mathcal{X}}_{AB_{L}} =  (\tilde{\mathcal{X}}_{A} \otimes \tilde{\mathcal{I}}_{B})$. \qed
\smallskip

Let us now address the case of \textit{imperfect bit commitment}. The setting is the same as above but now Bob has a non trivial probability of being able to distinguish the committed states~\cite{Dariano2007}. From Bob's perspective this defines an \textit{$\epsilon$-concealing protocol}, while it will necessarily enable Alice to define a \textit{$\delta$-cheating strategy}.

To formally define these two notions we introduce the trace distance between ontic distributions $\rho$ and $\sigma$, denoted  $D(\rho, \sigma) = 1/2 \sum_{i}^{4^{n_{b}}}|\rho_i - \rho_{\sigma_{i}}|$ where the index $i$ labels the $4^{n_{b}}$ diagonal entries of $\rho_{B}$. The trace distance equals $1$ for disjoint ontic distribution while it takes the value of $0$ for identical states. We also note that this distance is invariant under reversible transformations (i.e. permutations) as pairs of ontic states are permuted consistently.

In non perfect bit commitment  $D(\rho^{0}_{B}, \rho^{1}_{B}) \leq \epsilon$ and the protocol is known as \textit{$\epsilon$-concealing} protocol. When $\epsilon = 0$ the perfect case is recovered, however when $\epsilon > 0$ a new cheating strategy for Alice can be employed. We will here follow~\cite{Leverrier2014} and simplify the cheating strategy by letting Alice always commit to the state $\rho^{b = 0}_{AB}$ and consider her strategy to be the one which alters the value of $b$ to $1$. In particular, let the final state shared between $A$ and $B$ after the application of a dishonest strategy is given by $\sigma^{1}_{AB}$. In order to be successful, Alice will need to impose some bounds on the ability of Bob to discriminate between an honest and a dishonest strategy. Alice's $\delta$\textit{-cheating} strategy will fix a distance $D(\rho^{0}_{B},\rho^{1}_{B}) \leq \delta$ between the committed states, and a distance $D(\sigma^{1}_{AB},\rho^{1}_{AB}) \leq \delta$ on the revealed states, where $\sigma^{1}_{AB}$ is again the state obtained through the dishonest strategy. Such strategy is called a \textit{$\delta$-cheating} strategy.
The second inequality can be understood as defining a successful cheat for Alice as one where Bob cannot detect the difference between an honest Alice who encoded the bit value $1$ and the cheating Alice who changes it to that after the commitment phase. 

In order to prove imperfect bit commitment we will use the following lemma.

\begin{lemma}
Given two toy states $\rho_{B}$ and $\sigma_{B}$ such that their distance is $D(\rho_{B},\sigma_{B}) = \epsilon$, we can find two purifications $\psi_{AB}$ and $\phi_{AB} $ (of $\rho_{B}$ and $\sigma_{B}$ respectively) such that $D(\psi_{AB},\phi_{AB}) = D(\rho_{B},\sigma_{B}) = \epsilon $.
\label{thm:Bit_Commitment_Lemma}
\end{lemma}

\textit{Proof in Appendix~\eqref{App:Distance}}
\medskip

Following closely the strategy outlined in \cite{Leverrier2014} we now prove that given an $\epsilon$-concealing protocol there always exists an $\epsilon$-cheating strategy for Alice

\begin{theorem}
\textit{For any $\epsilon$-concealing toy bit commitment protocol there exists a valid $\epsilon$-cheating strategy}.
\label{thm:Imperfect_Bit_Commitment}
\end{theorem}

\textit{Proof.} Given the two reduced states $\rho^{0}_{B}$ and $ \rho^{1}_{B}$, there always exist two purifications $\psi_{AB}^{0}$ and $\phi_{AB}^{1}$ such that $D(\psi_{AB}^{0},\phi_{AB}^{1}) = D(\rho^{0}_{B},\rho^{1}_{B}) = \epsilon$ (see lemma \eqref{thm:Bit_Commitment_Lemma}). These two purifications can then be used to achieve an effective $\epsilon$-cheating strategy as follows. In virtue of Theorem \eqref{thm:Equivalence_Of_Purifications} there exists toy permutations $\tilde{\mathcal{U}} = \tilde{\mathcal{U}}_A \otimes \tilde{\mathcal{I}}_B $ and $\tilde{\mathcal{V}} = \tilde{\mathcal{V}}_A \otimes \tilde{\mathcal{I}}_B $ acting on Alice's system alone such that $\tilde{\mathcal{U}} \rho^{0}_{AB} \tilde{\mathcal{U}}^{T} = \psi_{AB}^{0} $ and $\tilde{\mathcal{V}} \rho^{1}_{AB} \tilde{\mathcal{V}}^{T} = \phi_{AB}^{1}$ respectively. 
We then define Alice's cheating strategy through $\sigma_{AB}^{1} := \tilde{\mathcal{V}}^{T}\tilde{\mathcal{U}} \rho^{0}_{AB} \tilde{\mathcal{U}}^{T} \tilde{\mathcal{V}} = \tilde{\mathcal{V}}^{T} \psi_{AB}^{0} \tilde{\mathcal{V}}$. Making use of the invariance of the distance, we thus obtain $D(\sigma_{AB}^{1}, \rho^{1}_{AB}) = D(\psi_{AB}^{0},\phi_{AB}^{1}) = D(\rho^{0}_{B},\rho^{1}_{B})\leq \epsilon$. \qed
\medskip

Theorem~\eqref{thm:Perfect_Bit_Commitment} proves Spekkens' conjecture in~\cite{Spekkens2007} that bit commitment is impossible in the toy model, while  theorem~\eqref{thm:Imperfect_Bit_Commitment} extends the result to the imperfect case. Further we note that theorem~\eqref{thm:Imperfect_Bit_Commitment} gives a different bound compared to the $\sqrt{2\epsilon}$-cheating strategy found for an $\epsilon$-concealing protocol in Gaussian quantum optics in~\cite{Leverrier2014}. The reason for this is that in the toy theory the two special purifications of lemma~\eqref{thm:Bit_Commitment_Lemma} have exactly the same distance as the reduced states Bob holds, while in the Gaussian case a weaker inequality is provided.

\subsection{Toy error correcting and secret sharing}
\label{sec:Error_Correction}

At first sight the nature of the toy model poses severe limitations on the kind of allowed toy error correction (EC) protocols. The toy model is an inherently classical theory, nonetheless featuring a no cloning theorem~\cite{Spekkens2007}: this implies that no EC protocols based on classical repetition codes can be employed. 
However, we here show that using the stabilizer structure of the toy model a valid EC toy protocol can be defined. In this case the translation from quantum to toy is less sensitive to choice of map for the generators.

We begin by providing an overview of the stabilizer quantum EC protocol~\cite{Gottesman1997}, as this will both explain the procedure and prepare the notation for the toy case. 
A quantum error correcting code takes $k$ qubits (we call these the `logical' qubits), and encodes them onto $n$ qubits (we call these the physical qubits, and $n>k$), in order to protect the information against noise. 
The key parameter of a code which describes its error correcting power is the distance $d$. A code of distance $d$ can correct for $\frac{d-1}{2}$ errors on unknown locations or $d-1$ errors on known locations (for example loss or erasure). Together we denote the code parameters as $[[n,k,d]]$. Our aim is to show that for any quantum stabilizer EC code $[[n,k,d]]$ there is a corresponding toy EC code with the same error correcting parameters.  We use single square brackets $[n,k,d]^{Toy}$ to parameterise  the toy EC in order to underline their classical nature.

One can picture the encoding as the space of the logical qubits (of dimension $2^k$) sitting inside the bigger space of the physical qubits (dimension $2^n$).
For a stabliser EC code~\cite{Gottesman1997}, the logical space is defined by a stablizer group, which we call the stabiliser code $ S_{C} = \langle g_{1}, \dots , g_{n-k} \rangle$.
For a particular encoding into this space, one can think of it as choosing a basis in this space.
Consider the pure stabiliser state $\rho_{S}$ defined on $k$ systems and described by stabilizer $S = \langle h_{1}, \dots, h_{k} \rangle $  and denote its encoding into the stabiliser code (the `logical' state) as  $S_{L}$. 
The generating set of $S_{L}$ can be decomposed into two distinct subgroups: 
The first is identified by the code's stabilizer group $ S_{C} = \langle g_{1}, \dots , g_{n-k} \rangle$, while the second subgroup $H_{L} = \langle h_{L_{1}}, \dots, h_{L_{k}} \rangle$ represents the logical generators of the particular encoding of the initial state $S = \langle h_{1}, \dots, h_{k} \rangle$. 
Explicitly, any encoded states is defined by stabilizer

\begin{equation}
\label{eq:EC_Encoding}
S_{L} = \langle g_{1}, \dots, g_{n-k}, h_{L_{1}}, \dots, h_{L_{k}} \rangle .
\end{equation}

The key to the translation to the toy case will be to notice that by writing the encoded state in the right way, errors essentially act trivially on it - and then that this also follows in the Toy version.
This is encapsulated in the following lemma. 

\begin{lemma}
Given a quantum EC code $[[n,k,d]]$, any of its logical encoded states admits a choice of generating set $S_{L} = <g_{1}, \dots, g_{n-k}, h_{L_{1}}^{\star}, \dots, h_{L_{k}}^{\star}>$ such that all its logical generators $h_{L_{1}}^{\star}, \dots, h_{L_{k}}^{\star}$ have simultaneously trivial support over up to $d-1$ arbitrarily chosen systems.
\label{thm:EC_Lemma}
\end{lemma}

\textit{Proof in appendix~\eqref{App:Proof_EC_Lemma}}.

\medskip

This lemma says that if the weight of an error is less than $d-1$ one can always write the logical part of the encoded state (in particular its generators), such that it is not touched by the error. As one follows through the workings of the code (see appendix~\ref{App:Error_Correction_Proof}), we see that this eventually means that it functions as a distance $d$ code.
In a sense the error simply shifts the `code space' of the encoding (by changing the code stabilisers), which are fixed back by the syndrome measurement and correction.

\medskip

To translate this to the toy framework now becomes very simple.  Given a toy state over $k$ systems we would like to protect, and an error $E$ such that $|E|\leq \frac{d-1}{2}$, one can directly translate the quantum case and see that the same reasoning can be applied. More details can be found in the appendix~\ref{App:Error_Correction_Proof}.
In this way a quantum code with parameters $[[n,k,d]]$ can be translated to a toy code $[n,k,d]^{Toy}$.

\begin{theorem}
\label{thm:Error_Correction_Theorem}
Given any $[[n,k,d]]$ quantum stabilizer error correcting code where $n$ is the number of physical systems, $k$ the number of the encoded systems, and $d$ is the code's distance, there exists a corresponding $[n,k,d]^{Toy}$ toy error correcting protocol which can correct any arbitrary toy error of weight $|E| \leq (d-1)/2$. 
\end{theorem}

\textit{Proof in appendix~\eqref{App:Error_Correction_Proof}}.

\medskip

We now briefly discuss an application of error correcting codes for secret sharing. 
In secret sharing, a dealer wishes to distribute a secret (say a bit, a qubit or a toy bit) to a network of players such that only authorised players can access the secret. The set of authorised sets is called the access structure.
Any secret sharing scheme can be loosely parameterised as a `ramp' scheme by three numbers - the number of players $n$, and two other numbers $l$ and $l'$, which denote that any set of $l$ or more players can access the secret, but no set of fewer than $l'$ can access any information at all about the secret. Together we call this an $(n, l, l')$ ramp secret sharing scheme. Note this parameterisation does not give any details about sets of size between $l$ and $l'$.

The connection between error correction and secret sharing was made in the quantum case first in \cite{Cleve1999}, where EC stabiliser codes are used to share quantum secrets and they show that all access structures  an be achieved that do not violate no-cloning. In~\cite{Marin2014} this was slightly refined using both the no-cloning and no-deletion theorem to connect the distance of a code and the ramp scheme parameters -  an $[[n,k,d]]$ error correction scheme functions as a ramp scheme with $l=n-d+1$ and $l'=n-l$, and vice versa.
It is not hard to see that these arguments follow through in the case of Toy error correction and sharing `Toy' secrets.

Firstly, one views secret sharing as an error code tolerating erasure - if a set of players of size $l$ can access the secret, this corresponds to being able to tolerate a loss of $n-l$ systems. In this way we see the distance must be at least $d=n-l+1$. Secondly, no-cloning implies that no two sets should be able to access the secret, so $l'\geq n-l$. Thirdly, when encoding onto pure states (as is the case for stabiliser error correcting code), the no-deletion theorem says the reverse -  if a set of players has no information, then its complement has all the information (obviously this can fail if one does not imagine the complement set to have the full purification of the encoding - hence the requirement for pure code states). This implies $l'\leq n-l$. Together this says that for pure state encoding $l'=n-l=d+1$.
As done in \cite{Cleve1999}, one can then use error correcting codes to give secret sharing schemes with the parameters required.

The existence of an EC scheme in the toy model can be interpreted as a statement about the way the information is encoded within a resource state. In particular, the existence of a toy EC and secret sharing protocols highlights the similarities within access structures and the way information is encoded between quantum and toy theory. It was already pointed out in \cite{Chiribella2011} the relationship between theories with purifications and the existence of error correction. Here, by exploiting the stabilizers structure of the toy model we further explicitly show how to construct a working error correction model and we recover its expected relationship to secret sharing.

From the point of view of Bell non-locality this result also shows that there are no genuine non-local effects at the heart of quantum error correction, nor of secret sharing. Indeed the existence of these protocols in the toy model implies that these properties are non-local in so far as they can be accounted by the steering properties of the model. 

\subsection{Measurement based toy computation}
\label{sec:MBTC}

Measurement based quantum computation (MBQC) is an universal scheme of quantum computation where computations are achieved by systematically performing single qubit projective measurements on an highly entangled resource state, known as a graph state \cite{Raussendorf2001, Raussendorf2013}. There are two main reasons behind the choice of defining an analogous scheme of computation in the toy model. Firstly, MBQC is based on graph states, a particular type of stabiliser states, which were defined in the toy model in~\cite{Pusey2012} and have already been implicitly used in the previous section on error correction and secret sharing. Secondly, is the existence of an MBQC based blind and verified computation protocol~\cite{Broadbent2008,Fitzsimons2012} which we are interested in translating to the toy model.

An MBQC computation begins by the creation of a graph state followed by a series of adaptive single system projective measurements which encode the computation~\cite{Raussendorf2001,Raussendorf2013,Danos2007,Mantri2016}. In order to counterbalance the randomness of quantum measurements, they are performed in layers and the measurements within each layer depend on (possibly all) the measurement outcomes of the previous layers~\cite{Browne}. 

The toy version, which we call measurement based toy computation (MBTC) will follow directly from the quantum case. Let us now describe in more details how an \textit{MBTC} computation is actually implemented. 


First, we define the graph state \cite{Pusey2012}. Consider a graph $G(V,E)$ composed of $V$ vertexes and $E$ edges. To each vertex is associated a single system, initially in state $\rho_{\mathcal{X}}$, i.e.  $S = \langle \mathcal{X}_{1}, \dots, \mathcal{X}_{n}\rangle $. Then for each edge between two vertexes a controlled $\tilde{\mathcal{Z}}$ permutation $\tilde{\mathcal{C}}_{Z}$ is performed. By using the subscripts to indicate the two systems, the transformation $\tilde{\mathcal{C}}_{Z}$ is defined as the operation that takes $ \mathcal{X}_{1} \otimes \mathcal{I}_{2} \rightarrow \mathcal{X}_{1}\otimes \mathcal{Z}_{2} $, $ \mathcal{I}_{1} \otimes \mathcal{X}_{2}\rightarrow \mathcal{Z}_{1} \otimes \mathcal{X}_{2} $, $ \mathcal{Z}_{1} \otimes \mathcal{I}_{2} \rightarrow \mathcal{Z}_{1} \otimes \mathcal{I}_{2} $, and $ \mathcal{I}_{1} \otimes \mathcal{Z}_{2}\rightarrow \mathcal{I}_{1} \otimes \mathcal{Z}_{2} $~\cite{Pusey2012}. 
The resulting state is called the graph state and has stabiliser group

\begin{equation}
S_{G} = \langle \mathcal K_{1}, \dots, \mathcal K_{n} \rangle, \text{ with } \mathcal K_{i} = \mathcal{X}_{i} \prod_{j \in N_{g}(i)} \mathcal{Z}_{j}, 
\end{equation}
where the $\mathcal K_{i}$ are the graph state generators, and $N_{g}(i)$ represents the neighbourhood of $i$. To speak about computations we identify two special subsets of the vertexes corresponding to the input $I \subseteq V$ and the output $O \subseteq V$. Computations are then driven by sequences of single systems measurements. If the output systems are also measured then a `classical' outcome is obtained, while if the outputs system remain unmeasured the output is encoded onto the state of the output systems. The pattern defined by these measurements together with the specifc graph resource and the input and output sets defines a specific measurement pattern (see \cite{Danos2007} for the quantum case).

As in quantum mechanics, measurements are not deterministic and hence the model must be equipped with a strategy to ensure the correcteness of the computation. This is achieved dividing the measurement pattern into layers and introducing a causal structure of corrections between them. 

In the quantum case it was shown in \cite{Danos2007,Browne} that  there exists a sufficient and necessary condition for the existence of a `correction' strategy, and it is provided by a graph-theoretical property known as generalized flow of the graph, or simply gflow. Informally, (g)flow exploits the stabilizer structure of the graph states by providing an algorithmic assignment of single system Pauli corrections to yet unmeasured systems whenever the $-1$ outcome for a measurement is obtained. The role of such operations is to correct the not yet measured  part of the resource into the state it would have been if the measurement had yielded the $+1$ outcome instead, for more details see \cite{Browne} and \cite{Markham2014}. Once the output layer of the computation is reached, either all the output systems are measured to obtain a classical output or they are kept in order to have a quantum output. 
 
We translate gflow to the toy case in appendix~\eqref{App:MBTC}, which similarly allows for determinstic operations to be carried out, forming the basis of MBTC. We then show MBTC to be capable of implementing an universal set of gates (permutation transformations). The obtained computational model is thus universal with respect to the computational class of the toy model. It is important to stress that the computational power of the toy model is indeed less than classical universal. We will analyze in greater details some consequence of this in section~\eqref{sec:Computational_Limitations}.

Furthermore, the families of graphs whose graph states are universal for MBQC are also universal for MBTC. This is true since graphs with measurements implementing all the cliffords in the quantum case are universal. These universal families of graphs, and the associated measurement patterns, are used in the construction of the protocols which follow.

\subsection{Blind and verified toy computation}
\label{sec:Blind_And_Verified}

An open question in computer science, with deep implications in physics (see, for example,~\cite{Aharonov2012}), is what is needed to verify a quantum computation. Imagine one had access to a quantum computer, how could it be demonstrated that the answers it provides are the correct ones for the classes of problems which are hard to simulate classically? Or even for the class of problems for which there is no polynomial time algorithm to check the correctness of an answer? And in particular, could a classical computer verify the correctness of these answers although it cannot compute the answer itself? Several protocols have been proposed during the past years to address such questions.  

While the general question of whether a full-scale quantum computer could be verified by a classical device is still open, the requirement on the verifier side have been reduced to some `small' degree of quantumness~\cite{Aharonov2012}. The quantity and the quality of this small degree of quantumness varies from protocol to protocol and, crucially to our exploration of toy protocols, it is not yet clear whether Bell non-locality is a necessary or simply sufficient requirement. While certain protocols (e.g.~\citep{Reichardt2013}) explicitly use Bell-non locality, the question is still open for others (e.g.~\cite{Fitzsimons2012}). 

Let us now provide a more concrete example of a interactive protocol while including the informal definitions of blindness and verification. These protocols are best understood in the framework of iterative proof systems, which corresponds to a delegated computation where a computationally weak trusted client (Alice, having the role of verifier) wants to delegate an hard, quantum computation to a more powerful but untrusted server (Bob, having the role of prover). The protocol must be such to satisfy the two following properties

\begin{enumerate}
	\item \textit{Blindness}: even if Bob is physically carrying out the computation for Alice, the input, the computation, and the output are at any given step always perfectly hidden to him. 
	\item \textit{Verification}: Alice can claim (with high probability) whether Bob has carried out the computation honestly according to the desired protocol or whether he has deviated from the honest computation.
\end{enumerate}

In other words, a blind and verified computation is one where Bob is carrying out the computation without gaining any information about it (blind) and if he tries to cheat he will be discovered (verified). Note that here verifiability does not imply directly a test of whether quantum mechanics is accurate, but it rather simply makes sure that the physical steps which compose the protocol have been faithfully performed by Bob~\cite{Aharonov2012}.

We now focus our attention on two protocols. The first one was introduced by Reichardt-Unger-Vazirani (RUV) in~\cite{Reichardt2013}, is based on the circuit model and makes explicit use of Bell non-locality in order to achieve the desired verification security. This protocol is therefore explicitly Bell non-local and hence impossible to be cast within the toy model. However there is another family of blind and verified protocols, those defined within the MBQC model introduced by Broadbent et al~\cite{Broadbent2008} and Fitzsimons and Kashefi (FK)~\cite{Fitzsimons2012}. 
Although not explicitly used, the question of whether Bell non-locality plays a role in this verification protocol is still open. What we are here interested in is better understanding whether Bell non-locality is \textit{necessary} in order to run the FK protocol or if correlations based only on the steering properties of quantum theory could be enough. 

We see that, indeed it is possilbe to define an analagous protocol for blind and verified computation. Following FK \cite{Fitzsimons2012} we first define a protocol for delegated computation, and then adapt it for blindness and verification.

\begin{figure*}
\includegraphics[width=\textwidth]{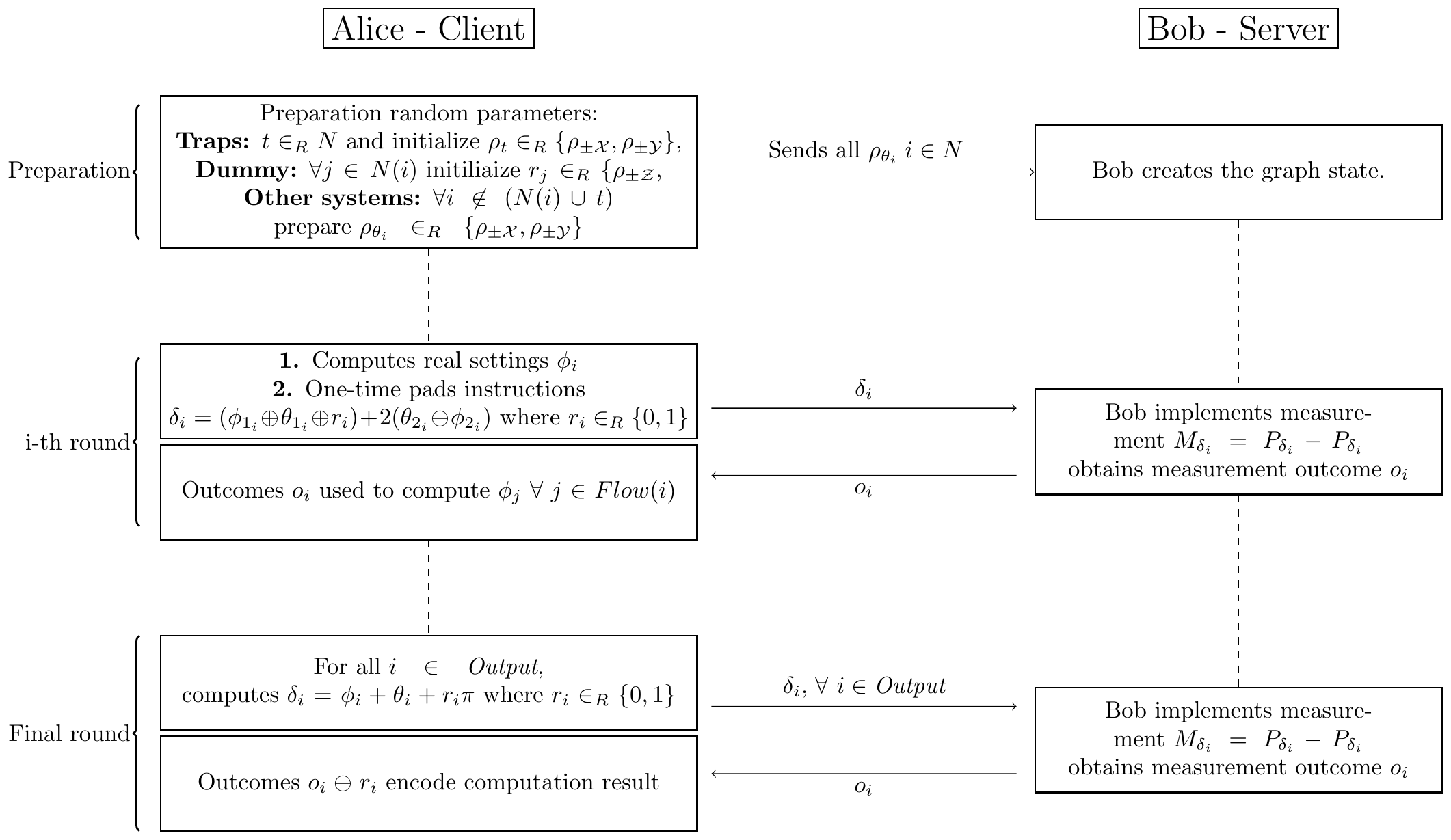}
\caption{\footnotesize Schematic view of the verified protocol assuming an honest Bob. First comes the preparation stage, where Alice chooses at random the location of the trap systems, and each system (trap, dummy, or computational) is one-time padded. The $i$-th round of the protocol is first composed by the computation of the gflow-corrected measurement settings $\phi_{i}$, followed by their encoding into the measurement settings $\delta_{i}$ sent from Alice to Bob. These measurement settings effectively undoes the state's one time padding $\theta_{i}$ and add an one-time pad on the outputs through $r_{i}$. Bob then performs a measurement $\{ P_{\delta_{i}}, P_{-\delta_{i}} \}$ and sends its outcome $o_{i}$ back to Alice. The protocol concludes on the last round when the systems $i \in O$, where $O$ is the output set, are reached. The outcomes for these measurements encode the output of the computation up to an $r_{i}$ bit flip.}
\label{fig:BlindVerifiedComputation}
\end{figure*}

\subsubsection{Delegated toy computation}

Let us now introduce the toy protocol for a delegated universal toy computation, which will form the basis for the blind and verified protocols to follow (see Figure~\eqref{fig:BlindVerifiedComputation}). Our construction  is fundamentally the same as the quantum one (see \cite{Broadbent2008} for more details), with some minor changes which will be pointed out in the next paragraphs. In practice the blind protocol will be built on top of the delegated one by randomizing certain parameters, while the verified protocol is constructed from the blind one with the addition of `traps'. 

The protocols are built using MBTC. One starts with a universal family of graphs, which is known to Alice and Bob (also for the blind and verified protocols). The choice of graph is made solely based on the input size. The gflow fixing, the causal order of measurements, is fixed by the graph. Finally the computation itself fixes the measurement angles. In this way, all the information, except the size of the computation (or rather an upper bound on this) is given by the measurement angles. These are what we will hide to make the protocol blind. 
First though, for a given input size the graph $G(V,E)$ is fixed, as is its gflow (known to Alice and Bob), and the delegated protocol is defined as follows:

\begin{defn} Delegated toy computation protocol
\begin{enumerate}
\item Alice prepares $N$ single toy systems, each system is initialized in the toy $\rho_{\mathcal{X}}$ state, and is sent to Bob.

\item Upon receiving the $N$ systems Bob entangles them into a toy graph state associated to graph $G(V,E)$.

\item \label{step: meas instructions} Alice sends to Bob the measurement instruction $\delta_i$ for the individual system $i$. The $i$-th system is chosen according to the sequence given by the gflow of the graph, starting with the first system $\delta_1$. Note $\delta_i$ is a classical message indicating which single system measurement $\{P_{\delta_i},P_{-\delta_i}\}$ should be performed.

\item Bob performs the measurement on system $i$, giving  measurement outcome $o_{i} \in \{0,1\}$, and sends $o_i$ to Alice.

\item Alice will use the measurement outcomes to calculate the measurement settings for the subsequent rounds of measurements labeled according to the gflow of the graph.

\item The protocol continues interactively until the last round of measurements is performed by returning to step \ref{step: meas instructions}. Finally the outcomes of the systems labeled as outputs encode the result of the computation.
\end{enumerate}
\label{defn:Toy_Blind_Protocol}
\end{defn}

\subsubsection{Blind MBTC}

In order to hide the computation and thus achieve blindness the above protocol needs to be modified in two key steps. \textit{First}, instead of preparing each state in a fixed basis, Alice initializes each system uniformly at random among a finite set of states. This will effectively one-time pad the initial resource state upon which the computation is performed. And \textit{second}, the initial mask is removed as Alice communicates Bob the measurement instructions: an extra bit of randomness is used in order to mix over the labels for the positive and negative outcomes of the measurement. 


To one-time pad the initial states, instead of always sending state $\rho_{\mathcal{X}}$ Alice chooses one of the four states in the set $A^{Toy} = \{\rho_{\mathcal{X}}, \rho_{\mathcal{Y}}, \rho_{-\mathcal{X}},\rho_{-\mathcal{Y}}\}$ uniformly at random. Here we see the first small difference between the toy and quantum protocols. In the quantum case, Alice chooses the initial states uniformly at random between the 8 states evenly spread on the equator of the Bloch sphere. The simple reason for the difference is that half of these are not stabiliser states, so we are left with Pauli states, which we use for our Toy case. This is related to the fact that the toy model is essentially as powerful only as Clifford quantum computation.

We next want to describe how Alice calculates the measurement basis, and how she injects the second randomisation through the random sign flip for the measurement basis. We start by fixing some notation. 
In order to keep notation as similar as possible to the original quantum protocol we will use lower case Greek letters $\delta,\phi,\theta$ to label both the choice of states and of measurement settings. 
The initial one time padding of Alice is represented by $\theta \in \{0,1,2,3\}$.
For convenience we will sometimes represent it by two bits $\theta_1,\theta_2 \in \{0,1\}$
\begin{equation}
\label{eq:Modulo_4}
\theta = \theta_{1} + 2\theta_{2} \in \{0,1,2,3 \}.
\end{equation}
Here $\theta_{1}$ can be understood as the bit fixing the sign of the state (measurement projector) and $\theta_{2}$ fixes the `basis' for the state (measurement projector).
In a slight abuse of notation we will use $\rho_\theta$ to denote the choice of the state in $A^{Toy}$ (so $\rho_{0}\equiv \rho_{\mathcal{X}}$, $\rho_{1}\equiv \rho_{\mathcal{Y}}$,$\rho_{2}\equiv \rho_{\mathcal{-X}}$,$\rho_{3}\equiv \rho_{\mathcal{-Y}}$ ). 

The choice of measurement setting is represented by $\delta\in \{0,1,2,3 \}$, corresponding to measuring operators $\{\mathcal{X}, \mathcal{Y}, -\mathcal{X}, -\mathcal{Y}\}$ respectively. 
The setting  depends on three parameters. Firstly, the angle of measurement as calculated by the gflow in the absence of any one-time padding $\phi \in \{0,1,2,3\}$ (this would be the measurement instructions in the delegated protocol of definition~\ref{defn:Toy_Blind_Protocol}, and it depends on previous measurement results). Again we write this in the form of two bits $\phi_1, \phi_2 \in \{0,1\}$ through $\phi = \phi_1 +2 \phi_2$ (similarly for $\delta$). 
Secondly we need to undo the one-time padding $\theta$. Finally, we add the final bit of randomness masking the measurement result using a bit $r\in \{0,1\}$. These are combined through
\begin{align}
\label{eq:toy_delta}
\delta &= \delta_{1} + 2 \delta_{2} \nonumber \\
&= (\phi_{1} \oplus \theta_{1} \oplus r) + 2(\theta_{2} \oplus  \phi_{2}) .
\end{align}

Note $r \in \{0,1\}$ changes the sign choice encoded in the setting $\delta$. This implies that when Bob performs the measurement associated to the projector $P_{\delta}$, the random variable $r$ will act as a one time pad over the sign of the projector hence randomizing its outcome. 


We are now ready to see that the above toy protocol~\eqref{defn:Toy_Blind_Protocol} satisfies two analogous properties of protocol $1$ in~\cite{Broadbent2008}: namely the above delegated toy protocol with the addition of the random parameters is blind (\textit{blindness}), and that these random parameter do not disrupt the computation (\textit{correctness}).

\begin{theorem} \label{thm:Blindness}
The the toy protocol~\ref{defn:Toy_Blind_Protocol} where each state is initially masked uniformly at random over $A^{Toy}$ and where each measurement setting is calculated according to equation~\ref{eq:toy_delta} is correct and blind.
\end{theorem}

\textit{Proof in appendix~\eqref{app:Blindness_proof}}.

\begin{figure*}
\includegraphics[width=\textwidth]{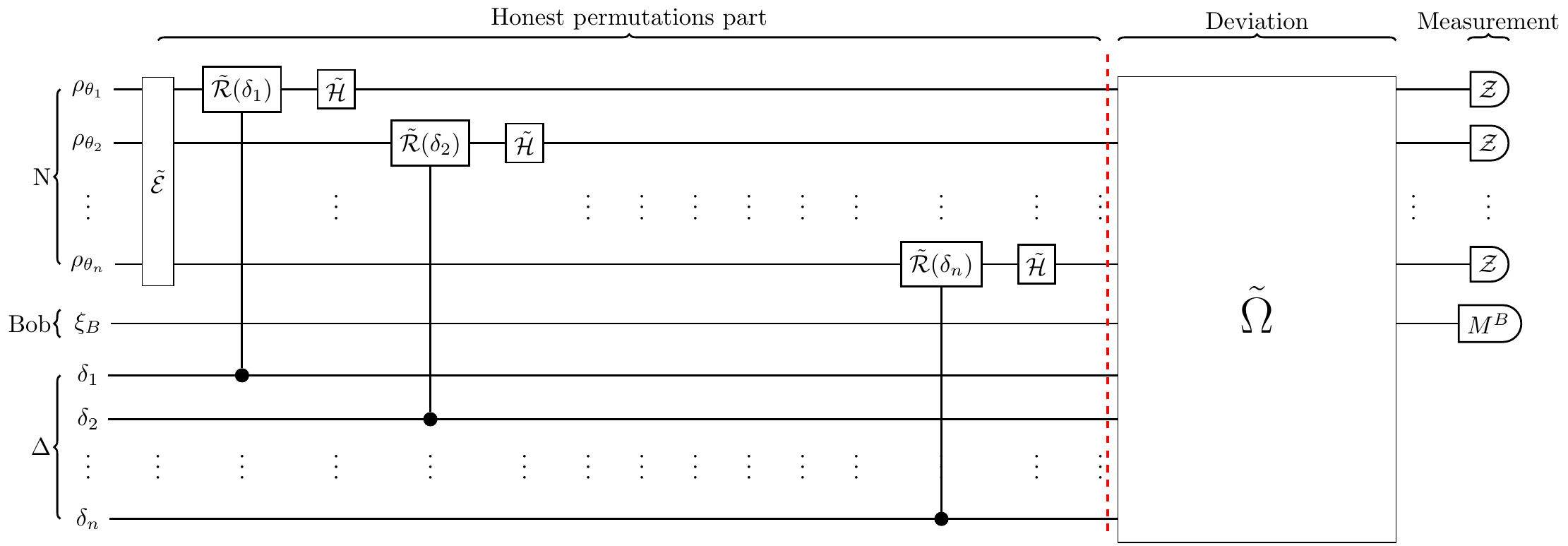}
\caption{\footnotesize The above circuit represents the decomposition of Bob's actions in the verification protocol. His actions are divided into the honest part of the computation (permutations) and a dishonest one (the deviation). There are three different sets of systems: the set $N$ contains the $n$ physical systems Alice sends to Bob; The set $Bob$ corresponds to Bob's private register; the set $\Delta$ represents the `classical' information encoding the measurement settings sent from Alice to Bob. An honest computation can be further divided into two parts: a permutation part and a measurement part. In the permutation part $\tilde{\mathcal{E}}$ creates the entanglement and is followed first by a controlled permutation $\tilde{\mathcal{R}}(\delta_{i})$ which removes the one-time pad $\theta_{i}$ from the state $\rho_{\theta_{i}}$ and then by a toy Hadamard $\tilde{\mathcal{H}}$. The measurement part then consists of single system $\mathcal{Z}$ measurements on all systems $i \in N$. However in order to account for Bob's deviation a global permutation over all the systems labelled $\tilde \Omega^{M,B,\Delta}$ is performed followed by some measurements $M^{B}$ on Bob private register. This describes Bob's most general deviation from the honest protocol.}
\label{fig:BobDeviation}
\end{figure*}

\subsubsection{Blind and verified MBTC}

We are now ready to translate the verification protocol. The protocol is itself a small modification of the blind scheme for delegated computation above. However in order to show that the figure of merit of the quantum protocol translates to the toy model it is necessary to explain many of its details.

We will now introduce the protocol (pointing out where it differs from the quantum one) and leave the full proof of its verifiability in appendix~\eqref{app:Verification_proof}, as the proof is rather techical and fudamentally a suitably adapted toy version of the original quantum one in reference~\cite{Fitzsimons2012}. Section~\eqref{sec:Computational_Limitations} will then feature a discussion on what the existence of a toy verified protocol means in light of the computational limitations of the toy model.

The key idea behind the verification protocol of~\cite{Fitzsimons2012} is the addition of trap qubits in the previous blind protocol which are used to test and unmask any possible deviation by Bob. Again we follow the quantum protocol exactly, with the limitation on the family of states and measurements. Informally, the trap systems are randomly initialized to a state $\rho_{t} \in_{R} A^{Toy}$ and carefully placed in the graph state by surrounding each trap by `dummy systems', which in turn correspond to systems randomly initialized in $  \mathcal{Z}$ or $ -\mathcal{Z}$. The dummy systems do not contribute to the computation but upon creating the desired resource state leave the traps systems disentangled from all the other non-traps systems. 
Therefore Alice, by knowing the true state of the trap system, deterministically knows its measurement outcome. Since Bob does not know the position of traps, or their true state,  the deterministic nature of the traps measurements offers a way to detect a malicious Bob. Alice then accepts the computation if the trap measurement results are correct and rejects if not.

The figure of merit of the protocol will be the probability of accepting as correct an incorrect computation. This will be called the probability of failure of the protocol, or $p_{fail}$. A formal definition of this quantity will appear in equation~\eqref{eq:Verification_Prob_Failiure}, as in order to do so we need to define all the terms that compose it. That is, label all the systems used in the protocol and specify the initial states, provide a definition of Bob's most general deviation from the honest protocol, clarify all the random parameters $\nu$, and define the projector onto an honest run of the computation $P_{honest}$ in order to be able to define its orthogonal projector $P_{\perp}$. 


Finally just as in~\cite{Fitzsimons2012} we will bound the probability of failure of the protocol when considering classical outputs (i.e. all systems are measured) and only one trap will be employed. A `circuit' representation of the protocol is given in figure~\eqref{fig:BobDeviation}.


\textit{First}, let us label the systems. Let $N$ and $\Delta$ label all the systems used to create the resource state and the measurement instructions respectively. Then select uniformly at random a position $t \in N$ for the trap, and proceed labelling the systems as follows. Let $O \subset N$ be the output systems, $I \subset N$ the input systems, and $D \subset N$ the dummy systems. The set of the dummies is completely specified by the location of the trap $t$. As mentioned, only `classical' output protocols will be considered which means that the systems labelled $O$ will be at the end measured and their outcomes define (after any necessary correction) the output of the computation. Finally, let B represents Bob's private auxiliary systems used in possible deviations. 

In the following the subscript indicates the system over which the state is defined. 
The state $\rho_{N \setminus t} \otimes \rho_{t} $ represents the initial one-time padded state defined over the computational, dummy, and traps systems (so $\rho_{N \setminus t}:= \otimes_{i\in N \setminus t}\rho_{\theta_i}$ and $\rho_{t} := \rho_{\theta_{t}}$);  $\delta_\Delta := \otimes_{i}^{N} \delta_{i}$ is the state containing the `classical' information encoding the measurement settings to be used for each system in $N$, that is including trap and dummy systems; and $\xi_{B} $ is the initial state of the auxiliary system in the hand of Bob. Bob's most general deviation can be defined as a generalized toy map (see~\eqref{sec:Generalized_Transformations}) composed by a global permutation $\tilde \Omega_{N,\Delta,B}$ followed by a partial trace over Bob's private subsystems $B$, where the support of the permutation is explicitly expressed in the subscripts.

\textit{Second}, let the following be all the random parameters used in the toy blind and verified protocol:

\begin{itemize}
	\item A position $t \in_{R} N$ for the trap. Note the trap can be placed \textit{anywhere} in $N$.	
	\item \sloppy A one-time pad $\theta_{t} \in_{R} \{0,1,2,3 \}$ for the trap's state $\rho_{\theta_{t}} $.
	\item  A one-time pad  $r_{t} \in_{R} \{ 0,1 \}$ to hide the measurement outcome of the trap.
	\item A one-time pad $\theta_{i} \in_{R} \{0,1,2,3\}$ for each system $\rho_{\theta_{i}} $ with $ i \in N \setminus (t, D)$.
	\item A one-time pad  $r_{i} \in_{R} \{ 0,1 \} $ to hide the measurement outcome for each system $i \in N \setminus (t, D)$.
	\item A choice $d_{i} \in_{R} \{ +1,-1 \}$ for the dummy states $  \rho_{d_{i}\mathcal{Z}_{i}} $. The position of the dummy systems $i \in D$ is fixed  by $t$ through $D = \{ j \in N | j \in N_{g}(t) \}$ and $N_{g}(t)$ identifies the neighborhood of the trap.
\end{itemize}

Importantly, all the parameters are chosen \textit{uniformly} at random. The parameters are further grouped with respect to whether they are defined for the traps or the computational and dummy systems. The trap's parameters are $\nu_{t} = \{t, \theta_{t}, r_{t}\}$, while the computational ones are $\nu_{c} = \{ \theta_{i},r_{i},d_{i}\}_{i \in N \setminus t}$. Regarding the dummy systems, note how the location of the $d_{i}$ depends only on the trap's position: once the location of a trap has been chosen all its neighbors become dummy systems and they are initialized to $ \rho_{d_{i}\mathcal{Z}_{i}}$ where $d_{i} \in_{R} \{ +1,-1 \}$ is chosen uniformly at random. It is important to stress that dummy systems \textit{do not} contribute the computation, they role is only to isolate the trap systems as they are not entangled by the $\tilde{\mathcal{C}}_{Z}$ operations which create the graph state. Further, note the outcome of either $\mathcal{X}$ or $\mathcal{Y}$ observable on the dummy state $\pm \mathcal{Z}$ has probability $1/2$. Finally, whenever $d_{i} = - 1$ the $\tilde{\mathcal{C}}_{Z}$ adds an extra $\tilde{\mathcal{Z}}$ to all vertexes neighboring $i$, this needs to be taken into account by Alice in order to decode the correct measurement outcomes provided by Bob.
\smallskip

\textit{Third}, we need to define the output of the computation both in the case Bob is honest and dishonest. To simplify the notation in what follows \textit{only} the non trivial support of states and operations will be explicitly labelled by a subscript. For example, a permutation written as $\tilde{\mathcal{C}}_{O}$ is a shorter version of $\tilde{\mathcal{C}}_{O} \otimes \tilde{\mathcal{I}}_{N \setminus O, \Delta}$. Let the output of the protocol be identified by the state $\rho^{out}_{O\setminus t, t}$, defined non trivially only on the output and trap systems. Since MBTC is adaptive as measurement settings depend via the gflow function on previous outcomes, we here give a circuit representation of the computation for a fixed vector $\vec{b}$ of measurement outcomes over $N\setminus(O,t)$ (see~\cite{Fitzsimons2012} for more details for the quantum case). In this way figure~\eqref{fig:BobDeviation} represents a possible run of the computation for a given list of outcomes $\vec{b}$, and that  each measurement setting $\delta_{i}$ depends on (possibly all) the previous measurement outcomes $j < i $, according to the gflow. Furthermore note the each MBTC measurement has been decomposed into a controlled $\tilde{\mathcal{R}}(\delta_{i})$ permutation and a toy $\tilde{\mathcal{H}}$ Hadamard, followed by a `computational basis' $\mathcal{Z}$ measurement. The controlled permutation is an allowed operation as it always amounts to a controlled toy Pauli permutation. The correct (i.e. honest) protocol before the measurements can thus be described by the permutation $\tilde{\mathcal{U}}^{hon}_{N,\Delta} = (\otimes_{i}^{N} \tilde{\mathcal{H}}_{i} \tilde{\mathcal{R}}_{i}(\delta_{i}))\tilde{\mathcal{E}}_{G}$, that is from right to left we have the creation of the resource state $\tilde E_{G}$ which consists of all the appropriate control-$\tilde{Z}$ operations, the controlled permutations from the $i$-th measurement setting to the $i$-th system in order to remove the one-time pad $\theta_{i}$, and finally the Hadamards to have $\mathcal{Z}$ measurements. In order to obtain the outcome of the protocol the average over all outcomes $\vec{b}$ will be taken.

In this way the output of the computation $\rho^{out}_{\Ot, t}$ is defined as a state that sits on the $\Ot$ output systems and on the trap $t$ (again note the trap can be placed \textit{anywhere}, and in particular, among the output systems),

\begin{align}
\label{eq:First_Output_state_in_full}
\rho^{out}_{\Ot, t} &= \sum_{\vec{b}} Tr_{(\N,\Delta,B)} [ \tilde{\mathcal{C}}_{\Ot}^{\bNu} P^{\vec{b}}_{\N} \tilde{\Omega}_{All} (\tilde{\mathcal{U}}^{hon}_{N,\Delta} ( \rho_{N\setminus t} \nonumber \\
&\otimes \rho_{t}^{r_{t}} \otimes \delta_{\Delta})\tilde{\mathcal{U}}^{hon^{T}}_{N,\Delta} \otimes \xi_{B}) \tilde{\Omega}^{T}_{All} P^{\vec{b}}_{\N} \tilde{\mathcal{C}}_{\Ot}^{\bNu}] \nonumber \\
&= \sum_{\vec{b}} Tr_{(\N,\Delta,B)} [ \tilde{\mathcal{C}}_{\Ot}^{\bNu} P^{\vec{b}}_{\N} \sigma_{All} P^{\vec{b}}_{\N} \tilde{\mathcal{C}}_{\Ot}^{\bNu}], 
\end{align}
\sloppy where the state $\sigma_{All} = \tilde{\Omega}_{All} (\tilde{\mathcal{U}}^{hon}_{N,\Delta} (\rho_{N\setminus t} \otimes \rho_{t}^{r_{t}} \otimes \delta_{\Delta}) \tilde{\mathcal{U}}^{hon^{T}}_{N,\Delta} \otimes \xi_{B}) \tilde{\Omega}^{T}_{All}$ is defined over all systems and depends on the honest computation through $\tilde{\mathcal{U}}^{hon}_{N,\Delta}$, on all the random parameters, on Bob's deviation $\tilde \Omega_{All}$, and on the specific measurement branch $\vec{b}$. 

In order to provide an explicit form of the output state consider performing the measurements over the $\N$ systems. These are all $\mathcal{Z}$ measurements and their outcomes are encoded in the vector $\vec{b}$ which implies that after the projections are taken into account the state becomes 

\begin{align}
\rho^{out}_{\Ot, t} =  \sum_{\vec{b}} Tr_{(\N,\Delta,B)} &[ p(\vec{b}) \rho^{\vec{b}}_{\N}  \\
&\otimes \tilde{\mathcal{C}}_{\Ot}^{\bNu}  (\sigma_{\Ot,t,\Delta,B}^{\vec{b}}) \tilde{\mathcal{C}}_{\Ot}^{\bNu}] \nonumber,
\end{align}
where $p(\vec{b}) = Tr[P^{\vec{b}}_{\N} \sigma_{All} P^{\vec{b}}_{\N}] $ is the probability of obtaining the outcomes $\vec{b}$ for the $\N$ measurements, $\rho^{\vec{b}}_{\N}$ is the post-measurement state over $\N$ with stabiliser group $ \langle \nobreak b_{1} \mathcal{Z}_{1}, \dots, b_{\N} \mathcal{Z}_{\N} \rangle$, and $\sigma_{\Ot,t,\Delta,B}^{\vec{b}}$ is the remaining part of the state $\sigma_{All}$ which now depends on the particular measurement branch $\vec{b}$. This post-measurement product state $\rho^{\vec{b}}_{\N} \otimes \sigma_{\Ot,t,\Delta,B}^{\vec{b}}$ is obtained via the usual update rules (see section~\eqref{sec:Measurements}). Then the partial traces over $\Delta$ and $B$ can be taken giving 

\begin{equation}
\label{eq:Verification_Output_Computation}
\rho^{out}_{\Ot, t} = \sum_{\vec{b}} p(\vec{b}) Tr_{\N} [ \rho^{\vec{b}}_{\N} \otimes \tilde{\mathcal{C}}_{\Ot}^{\bNu}  (\sigma_{{\Ot,t}}^{\prime \vec{b}}) \tilde{\mathcal{C}}_{\Ot}^{\bNu}],
\end{equation} 
where $\sigma_{{\Ot,t}}^{\prime \vec{b}} = Tr_{(\Delta,B)} (\sigma_{\Ot,t,\Delta,B}^{\vec{b}} )$. This expression defines the output of the protocol.

Now consider the case of honest run of the protocol: in such a scenario $\tilde \Omega = \tilde{\mathcal{I}}$ and performing the traces over the $\Delta$ and $B$ systems yields

\begin{equation}
\rho^{honest}_{\Ot, t} = \sum_{\vec{b}} Tr_{\N} [ p(\vec{b}) \rho^{\vec{b}}_{\N} \otimes \tilde{\mathcal{C}}_{\Ot}^{\bNu} \rho^{\vec{b}}_{\Ot} \tilde{\mathcal{C}}_{\Ot}^{\bNu} \otimes \rho_{t}^{r_{t}}]. \nonumber
\end{equation}

Note that this is now an honest run of the protocol, which firstly implies that the trap systems and the $\Ot$ ones are in a product state, and secondly that the correction acting over the $\Ot$ systems always yields the correct output state of the computation: $\rho^{honest}_{\Ot} = \tilde{\mathcal{C}}_{\Ot}^{\bNu} \rho^{\vec{b}}_{\Ot} \tilde{\mathcal{C}}_{\Ot}^{\bNu}$. Therefore

\begin{align}
\label{eq:Verification_Honest_Computation}
\rho^{honest}_{\Ot, t} &= \sum_{\vec{b}} Tr_{\N} [ p(\vec{b}) \rho^{\vec{b}}_{\N} \otimes \rho^{honest}_{\Ot} \otimes \rho_{t}^{r_{t}}] \nonumber \\
&= \rho^{honest}_{\Ot} \otimes \rho_{t}^{r_{t}},
\end{align}
where the last line is obtained first by summing over all the outcomes $\vec{b}$ and, noticing that $\sum_{\vec{b}} p(\vec{b}) = 1$, performing the trace over $\N$. Having thus defined an honest computation, let $P^{honest}_{\Ot}$ be the projector associated to the honest state $\rho^{honest}_{\Ot}$. The \textit{figure of merit} of the toy protocol is formally defined as the probability of failure:

\begin{equation}
\label{eq:Verification_Prob_Failiure}
p_{fail} = \sum_{t,r_{t},\nu_{c}} p(t,r_{t},\nu_{c}) Tr_{O \setminus t, t} \big[[( I_{O \setminus t} - P^{honest}_{O \setminus t})\otimes P^{r_{t}}_{t}] \rho^{out}_{\Ot,t} \big],
\end{equation}
here $\rho^{out}_{\Ot,t}$ is the state of equation~\eqref{eq:Verification_Output_Computation}, $p(t,r_{t},\nu_{c})$ are the probabilities associated to each random variable, and $P_{t}^{r_{t}}$ is the projector onto the correct traps state. The above toy protocol can then be defined to be $\epsilon$-secure if its probability of failure goes as $p_{fail}= 1 - \epsilon$ with $\epsilon > 0$.

\begin{theorem}
\label{thm:Verification_Theorem}
The toy version of the FK protocol with a single trap is at least $1 - \frac{1}{2n}$ secure, where $n$ is the total number of systems employed.
\end{theorem}

\textit{Proof in appendix~\eqref{app:Verification_proof}}.
\medskip

Hence we provide a blind and verified toy protocol based on a slight variation of the quantum FK protocol. This result strongly suggest that steering properties seem indeed to be sufficient to implement the FK protocol, even in the quantum setting. As in the toy verification fundamentally a computationally bounded \textit{classical} machine verifies an \textit{unbounded} machine an objection could be raised as maybe some form of Bell non locality may be required when full universal quantum computation is tried to be verified. Further we note that recent results in~\cite{Gheorghiu2015} indicate that the security of the FK protocol can be built on the steering properties of quantum theory. 

It is also important to note that the difference in computational power for our toy protocol is somewhat weaker than in the quantum case (assuming quantum computers are more powerful), but the main interest we believe is in the ability to verify in itself. In our case Alice and Bob both have access to classical computers, however Alice is only able to do computations of a limited size (constant), while Bob's computational power is unbounded. 

\subsection{Computational limitations}
\label{sec:Computational_Limitations}

After having analyzed all these toy protocols it is important to make a few clarifications about both the computational power and the `usefulness' of the toy theory as a realistic physical model of computation. Here, in line with Spekkens' original paper, rather than taking the toy theory as a realistic model for information processing tasks we are interested in seeing which quantum protocols can be successfully implemented within the toy model as this will give a concrete example of a protocol that can be cast in a Bell non-local theory.

Let us start by recalling that it was already shown in \cite{Spekkens2007} and \cite{Pusey2012} the simulation of the toy model is complete for the classical complexity class $\oplus L$ known as `parity L'. The class $\oplus L$ can be described quantumly as the class of problems which can be reduced to a polynomial size circuit consisting of only NOT and Control-NOT gates acting on the quantum state $\ket{00\dots 0}$ \cite{Aaronson2004}. More colloquially it fundamentally limits the computational power of the toy model to something weaker than even universal classical computation. Therefore there is no hope of reproducing any computational speed-up in the toy model~\cite{Spekkens2007,Pusey2012}. However, quantum information processes feature many other interesting non-computational advantages, and we are here interested in studying whether these kind of protocols can be run within the Spekkens' toy model.


We can therefore group our results into two distinct categories. On the one hand are the protocols which are not focused on computations such as bit-commitment, error correction, and secret sharing. The existence of such protocols in the toy model implies that the resource needed in the quantum case need not be based on Bell non-locality. However, on the other hand we have the computational protocols of blind and verified MBTC. Here there are two important considerations to be made. In order to define these protocol we moved outside the initial domain of the toy model by allowing the toy model to perform the modulo $4$ operations needed do decrypt the states in the blind protocol (see equation \eqref{eq:toy_delta}), and the power of creating random states (Alices one time padded initial states). These two features can potentially upgrade the power of a toy computer: the ability of performing modulo $4$ operation gives the toy model the ability of evaluating non-linear functions hence upgrading its computational power to classical universal~\cite{NielsenMichaelA.andChuang2010}, and the power to create random state can, in principle, create distributions which are not contained within the standard toy model of~\cite{Spekkens2007,Pusey2012}. 



Consider first the use of modulo $4$ computation. This is necessary for the blind protocol of~\cite{Broadbent2008}. It comes about because of the choice from one of four states and measurement settings. Without this hiding it is not clear that Alice keeps enough information hidden from Bob, and as such it is integral part of the protocol. 
In order to nonetheless define the toy blind protocol we allow the toy model to have access to a classical resource which implements the decoding operation, any time it might be needed at the price of further restricting the use of this modulo $4$ resource only to the decryption stage. That is, we step outside the boundary of the toy model allowing it to interact with a machine which only performs the required modulo $4$ limited only to the decryption stage of the computation. This way the non-toy operation is limited to some \textit{external} processing to the actual toy computation.

Now  we turn our attention the use of randomness in the toy model. We note a toy model embedded in a toy world is somewhat inconsistent when it comes to creating states ex-novo as the simple act of having to decide in which state to create a single elementary system requires the ability to choose between the 6 toy pure state and the maximally  mixed one, and this computation already exceeds the power of a linear machine such as the toy model. We here  take the view that Alice holds a black box which provides uniformly at random any of the $6$ pure toy states, along with a label indicating which state has been prepared. Furthermore, we could allow Alice to have access to an $n$ sided coin, as she will to pick the position $t \in_{R} N$ for the trap. Note however that in general this may allow the generation of ontic distributions which sit outside the toy model. Therefore we here pick the view that, like for each state initialization, Alice's black boxes will be such that they attach a label `trap' with probability $1/n$ to each prepared system. From Alice's point of view this corresponds to an allowed state, while from Bob's perspective this will correspond to an equally weighted mixture with respect to all random parameters. The fact that from Bob's point of view this is not a valid state is an harmless consequence of blindness, since his description of the state is yes statistical, but does not correspond to any real epistemic toy state.

These two modification are necessary in order be able to translate the protocol of~\cite{Fitzsimons2012}. However, by limiting their use to very specific situations we do not destroy the local essence of the toy model and, importantly, these additions are not enough to introduce Bell non-locality in the Spekkens' toy model.

Furthermore these extensions appear more reasonable when considering related `toy' models such as the subset of quantum theory given by Gaussian quantum optics. If we were provided with a Gaussian source there is no question that the only states that the source will be able to prepare are Gaussian. In this scenario the use of a $n$ sided coin to chose which state to create among the set of available Gaussian states may force the description of a state by someone who is not given access to the coin's as non Gaussian. However, this would not imply that the source is actually producing non-Gaussian states: the randomness is not \textit{physically} used to create states outside the theory, but rather is used to mask the description of the state by someone who has no access to the coin. Further, the ability of modulo $4$ addition can also be added to the Gaussian example with less conceptual difficulties. The reason being that Gaussian quantum optics is embedded within our world, which is not `toy', while previously we were fundamentally trying to consider the toy model living in a toy world.

\section{Conclusions \& Further work}
\label{sec:Conclusion}

In this work we have seen that a particular non-contextual, epistemic theory supports several protocols with analogies from quantum information. We know however that it cannot support all protocols - in particular any that relies on Bell non-locality, such as those providing device independent security. A natural question is where does this boundary lie. If someone presents a new protocol tomorrow, what guidance can we take from the work here about the existence of analogies in a toy theory?

Computationally there is evidence that contextuality is required for any computational speed up - which is consistent with our results here (our model is computationally trivial)~\cite{Howard2014} - so we do not expect speed up.
On the other hand, it seems from our results that the way in which information is spread and accessed is analogous to quantum theory - as we see for error correction and secret sharing.

More generally for cryptographic applications, we would argue that our results suggest the toy theory can provide security in cases where one party is allowed to make assumptions about the origins of correlated statistics.
In the case of the verification protocol presented here Alice trusts her preparation of states. In other cases, such as one-sided device independent quantum key distribution (QKD) she trusts her measurement devices  - the connection to steering here, and the hierarchy of trust and correlations in QKD is presented elegantly here~\cite{Branciard2012}. 
We can then anticipate for example that one sided device independent QKD should also work in the toy theory.


In the other direction, considering those protocols which quantum mechanics prohibits, we have seen that toy bit commitment is impossible as in the quantum case, confirming the conjecture by Spekkens~\cite{Spekkens2007}. More generally this can be understood as a consequence of the no-deletion property of the toy theory, adding to the list of analogous properties of the theory.

In recent years, the ability to process information in different ways has become a way of characterising and separating classical from quantum. Our results demonstrate further some of the subtleties involved in what we call classical. 
On the other hand our results may also be of practical interest. 
Several related toy theories can be viewed as practically motivated restrictions of quantum mechanics, in particular Gaussian quantum optics~\cite{Bartlett2012}.
Recognising where or which protocols can suggest where they may be implemented with greater ease.
\smallskip

\begin{acknowledgments}
The authors would like to thank Elham Kashefi, Theodoros Kapourniotis, and Tommaso Demarie for useful discussions and feedback. We further acknowledge support from the ANR project COMB and the ville de Paris project CiQWii.
\end{acknowledgments}

\appendix
\section{A simple toy exampe} \label{App:AppendixA}

As an example of some basic manipulations of toy states, consider a single system described by the epistemic state $1 \vee 2$. This state is equally describable as the $+1$ eigenstate of the toy $\mathcal{Z}=diag(1,1,-1,-1)$ observable. That is, $\langle \mathcal{Z} \rangle=\{Z, \mathcal{I}\}$ is the stabilizer group for the epistemic state $1 \vee 2$. Finally we define the state matrix 

\begin{equation*}
\rho_{\mathcal{Z}} = \frac{|S|}{4} P_{\mathcal{Z}} = \frac{1}{4}(\mathcal{I} + \mathcal{Z}) = \frac{1}{2}
\left( \begin{array}{cccc}
1 & 0 & 0 & 0 \\
0 & 1 & 0 & 0 \\
0 & 0 & 0 & 0 \\
0 & 0 & 0 & 0 \end{array} \right).
\end{equation*}
We see that  $Tr(\mathcal{Z} \rho_{\mathcal{Z}}) = 1$. On the other hand, the orthogonal state to $1 \vee 2$ is identified by $3 \vee 4$ or equivalently $\langle -\mathcal{Z} \rangle$. Again, we can also write the ontic distribution and the projector as follows

\begin{equation*}
\rho_{-\mathcal{Z}} = \frac{|S|}{4} P_{\mathcal{Z}} = \frac{1}{4}(\mathcal{I} - \mathcal{Z}) = \frac{1}{2}
\left( \begin{array}{cccc}
0 & 0 & 0 & 0 \\
0 & 0 & 0 & 0 \\
0 & 0 & 1 & 0 \\
0 & 0 & 0 & 1 \end{array} \right).
\end{equation*}

A permutation on the state gives another state

\begin{equation*}
\rho_{g} = \tilde{\mathcal{U}} \rho_{\mathcal{Z}} \tilde{\mathcal{U}}^{T},
\end{equation*}

where $g \in G_{1}$ and $\tilde{\mathcal{U}} \in U_{1}$ is a single system permutation. 
Consider measurements of the toy observables $\mathcal{Z}$ and recalling that identity is the only observable with not vanishing trace

\begin{align*}
Tr(P_{\mathcal{Z}} \rho_{\mathcal{Z}}) &=  \frac{1}{2\times 4} Tr((\mathcal{I} + \mathcal{Z}) (\mathcal{I} + \mathcal{Z})) \\ &= \frac{ Tr(\mathcal{I})}{4} = 1, \\
Tr(P_{-\mathcal{Z}} \rho_{\mathcal{Z}}) &=  \frac{1}{2\times 4} Tr((\mathcal{I} + \mathcal{Z}) (\mathcal{I} - \mathcal{Z}))  \nonumber \\ &= \frac{Tr(\mathcal{I}) - Tr(\mathcal{I})}{8} = 0. \nonumber
\end{align*}
That is, the probability of obtaining outcome $+1$ for a $\mathcal{Z}$ measurement on the state $\rho_{\mathcal{Z}}$ is unity, while the probability of obtaining $-1$  is zero. Similarly we can check the probability of obtain results $\pm 1$ for the $\mathcal{X}$ measurement we obtain

\begin{equation*}
Tr(P_{\pm \mathcal{X}} \rho_{\mathcal{Z}}) =  \frac{1}{4} Tr( (\mathcal{I} \pm \mathcal{X}) (\mathcal{I} + \mathcal{Z})) = \frac{1}{2},
\end{equation*}

which means that with equal probability we will either obtain $+1$ or $-1$ to such measurement. The new state after the measurement will then be $S = \langle \pm \mathcal{X} \rangle$ according to the measurement outcome. Observe how the two measurement are non compatible as measuring $\mathcal{X}$ before $\mathcal{Z}$ is not in general the same as measuring $\mathcal{Z}$ before $\mathcal{X}$.

\section{Distance of purifications} \label{App:Distance}

When writing a bipartite state $\psi_{AB}$ we can employ a useful graphical representation as follows. 
The $4^{n_{a } + n_{b}}$ diagonal entries of $\psi_{AB}$ can be seen as belonging to a $4^{n_{a}} \times 4^{n_{b}}$ grid where the rows label the $4^{n_{a}}$ ontic states belonging to $A$ and similarly the columns identify the ontic states of $B$. 
To do so we  explicitly assign a value to each pairs of ontic states between the two systems by $\psi_{AB_{(i,j)}}$, where $i \in \{ 1, \dots, 4^{n_{a}} \}$ labels the $4^{n_{a}}$ ontic states belonging to $A$ and similarly  $j \in \{ 1, \dots, 4^{n_{b}} \}$ for $B$. To simplify the analysis of what follows we again take, without loss of generality, $n_{a} = n_{b} = n$. Further since $\psi_{AB}$ is pure $rank(\psi_{AB}) = 4^{n}$. For the reader familiar with Spekkens' original paper the $i,j$ labeling is fundamentally describing any bipartite system as a pair of elementary systems. We now need to show that for any two reduced states $\rho_{B}$ and $\sigma_{B}$ such that their distance is $D(\rho_{B},\sigma_{B}) = \epsilon$ we can find two purifications $\psi_{AB}$, such that $\rho_{B} = Tr_{A}(\psi_{AB})$, and $\phi_{AB} $, such that $\sigma_{B} = Tr_{A}(\phi_{AB})$, defined in order have distance $D(\psi_{AB},\phi_{AB}) = D(\rho_{B},\sigma_{B}) = \epsilon $. Note how due to the triangle inequality any pair of purifications always has an equal or greater distance than that of the original (reduced) states:

\begin{align}
D (\rho_{B}, \sigma_{B}) &= \frac{1}{2} \sum_{i = 1}^{4^{n_{a}}}  |\rho_{i} - \sigma_{i}| \\
&= \frac{1}{2} \sum_{i = 1}^{4^{n_{a}}} | \sum_{j = 1}^{4^{n_{b}}}\psi_{i,j} - \sum_{j = 1}^{4^{n_{b}}} \phi_{i,j}| \\
&= \frac{1}{2} \sum_{i = 1}^{4^{n_{a}}} | \sum_{j = 1}^{4^{n_{b}}} (\psi_{i,j} - \phi_{i,j})|  \\
&\leq \frac{1}{2} \sum_{i = 1}^{4^{n_{a}}}\sum_{j = 1}^{4^{n_{b}}} | \psi_{i,j} - \phi_{i,j}| .
\end{align}

All is left is to find two explicit purifications such that they saturate the inequality. Note the triangle inequality is saturated only when, for each row $i$ independently, the terms $(\psi_{i,j} - \phi_{i,j})$ are all either $(\psi_{i,j} - \phi_{i,j}) \geq 0$ or $(\psi_{i,j} - \phi_{i,j}) \leq 0$. 

For our purposes we can take one of the reduced state, say $\rho_{B}$, to always represent the maximally mixed state over the partition $B$. That is, $\rho_{B} = \mathcal{I}_{\otimes n}$. Then we can show that regardless the form of $\sigma_{B}$ we can always find two purifications $\psi_{AB}$ and $\phi_{AB}$ such that they always saturate the above bound. 
Since $\rho_{B} = \mathcal{I}_{\otimes n}$ then by construction $\psi_{AB}$ has to be a maximally mixed state. For simplicity take the purification $\psi_{AB}$ to be defined as

\begin{equation}
\psi_{i,j} = \begin{cases} 
 \frac{1}{4^{n}} \text{ when } i = j, \\
 0 \; \text{ when } i \not = j.
\end{cases}
\end{equation}

That is, in the graphical representation $\psi_{AB}$ corresponds to a grid where the diagonal from bottom left to top right are all shaded. Any other choice of maximally correlated state will simply correspond to a permutation of its ontic states. Take then the other system $\sigma_{A}$ to be any arbitrary toy system. Its diagonal entries form a vector whose entries are labeled as $\sigma_{A_{i}}$ where $i \in \{1, \dots, 4^{n}\}$. We then offer a systematic way to construct a purification $\phi_{AB_{(i,j)}}$ of this state such that it saturates the triangle inequality. Whenever $\sigma_{i} =0$ then all entries $\phi_{AB_{(i,j)}}  = 0$ for all $j$, because all entries are positive and need to sum up to zero. In this case all the terms $(\psi_{i,j} - \phi_{i,j}) \geq 0$. Conversely for all $i$ where $\sigma_{i} \not = 0$ then always select the purification to have $\phi_{AB_{(i,j)}} = \frac{1}{4^{n}}$ and for all the other entries $j \not = i$ selects values of $\phi_{AB_{(i,j)}}$ such that $\sigma_{A_{i}} = \sum_{j}\phi_{AB_{(i,j)}}$ and that they form an over all valid state. Then for all such cases we obtain $(\psi_{i,j} - \phi_{i,j}) \leq 0$. Therefore this construction creates a  $\phi_{AB}$ which always saturates the inequality.

\section{Proof Lemma 6} \label{App:Proof_EC_Lemma}

Take a $[[n,k,d]]$ quantum EC code, an arbitrary logical state $S_{L} = \langle g_{1}, \dots, g_{n-k}, h_{L_{1}}, \dots, h_{L_{k}} \rangle$, and the \textit{erasure} error, described by the loss of $d-1$ systems denoted by $D$ ($|D| = d-1$) . By definition of the code, this is a correctable error and it can be described by the partial trace operation over $D$, followed by an initialization of the lost systems to identities. The resulting post-error state is then stabilized by

\begin{equation}
\label{eq:Lemma_Ec_Noisy_state}
S_{Tr} = \{ s_{i} | s_{i} \in S_{L} \text{ and } s_{i} = \mathcal{I_D}\otimes a_{D^C} \}.
\end{equation}

That is, the surviving stabilisers are idenity over $D$. Upon performing the $n-k$ syndrome measurements on the state stabilized by $S_{Tr}$ the new stabilizer

\begin{equation}
S^{\prime} = \langle \mu_{1} g_{1}, \dots, \mu_{n-k} g_{n-k}, s_{1}, \dots, s_{k} \rangle
\end{equation}
is obtained where each $\mu_{i} \in \{\pm 1 \}$ and a recovery operation $R^{\vec \mu}$ is identified. Note that there are $k$ linearly independent $s_{j}$ elements and  $s_{j} \in S_{Tr}$. This is true since all elements of $S_{Tr}$ commute with all the $g_i$, as we also have $s_{j} \in S_{L}$. Furthermore, as the error is correctable we know that $R^{\vec \mu^{\dagger}} (\rho_{S^{\prime}}) R^{\vec \mu} = \rho_{S_{L}}$. As the recovery is an unitary operation it suffices to consider its action with respect to the generating set. By construction we have that the code's generators satisfy

\begin{equation}
R^{\vec \mu^{\dagger}} \mu_{i} g_{i} R^{\vec \mu} = g_{i}, \quad \forall \; g_{i} \in S_{L},
\end{equation}
and this implies that by an appropriate labeling and choice of $s_i$

\begin{equation}
R^{\vec \mu^{\dagger}} s_{i} R^{\vec \mu} = \prod_{j\in A} g_{j} h_{L_{i}},  
\end{equation}
for each of the logical generators $h_{L_{i}}$, for some set $A$. Inverting the relation and using the fact that $R^{\vec \mu}$ is a Pauli operation, we have $s_i = R^{\vec \mu}  \prod_{j} g_{j} h_{L_{i}} R^{\vec \mu^{\dagger}} = \alpha(\vec \mu ) \prod_{j} g_{j} h_{L_{i}}$ where $\alpha(\vec \mu ) \in \{+1,-1\}$ depending on the commutation between the recovery operation and the given stabilizers. However, since $s_i \in S_L$ and clearly $g_{j} h_{L_{i}} \in S_L$ we have $ \alpha(\vec \mu ) =1$. Therefore, for some set of code generators $A$ we have

\begin{equation}
s_{i} =  \prod_{j\in A} g_{j} h_{L_{i}}, \quad \forall \; i.
\end{equation}

By definition $s_i \in S_{Tr}$ therefore is trivial over systems $D$. Defining then $h_{L_{i}}^{\star} = \prod_{j} g_{j} h_{L_{i}}$ concludes the proof. \qed 

\section{Proof of Theorem \ref{thm:Error_Correction_Theorem} (Error Correction)} \label{App:Error_Correction_Proof}

We now begin by tracing through an error correction process using the quantum stabiliser EC codes~\cite{Gottesman1997}, starting with the encoding, followed by some error, followed by syndrome measurement and correction. We take a stabiliser state to be encoded, as this is what we are interested in seeing in the Toy case, but the discussion can easily be adapted to more general quantum states.

We start by taking a $[[n,k,d]]$ stabiliser EC code.
A pure stabiliser state $\rho_{S}$ defined on $k$ systems and described by stabilizer $S = \langle h_{1}, \dots, h_{k} \rangle $ be encoded within a larger collection of $n > k$ systems. 
Let $S_{L}$ be the stabilizer of the corresponding logical states. 
It is important to notice the generating set of $S_{L}$ can be decomposed into two distinct subgroups: The first is identified by the so called code's stabilizer $ S_{C} = \langle g_{1}, \dots , g_{n-k} \rangle$, where each $g_{i}$ is colloquially called a code generator, while the second subgroup $H_{L} = \langle h_{L_{1}}, \dots, h_{L_{k}} \rangle$ represents the logical generators of the particular encoding of the initial state $S = \langle h_{1}, \dots, h_{k} \rangle$. Explicitly, any of the $2^{k}$ encoded states is defined by stabilizer

\begin{equation}
\label{eq:EC_Encoding}
S_{L} = \langle g_{1}, \dots, g_{n-k}, h_{L_{1}}, \dots, h_{L_{k}} \rangle .
\end{equation}

Consider now an error acing on systems $E$, such that $|E|\leq\frac{d-1}{2}$, described by  a noise map $\mathcal{F}$ which in its most general form, in the quantum case, corresponds to a CPTP map. For our purposes we will only consider errors $\mathcal{F}$ such that they produce other stabilizer states. 
By Lemma~\eqref{thm:EC_Lemma}, one can define an equivalent set of logical operators $h_{L_{i}}^\star=\prod_{j \in A_i} g_j h_{L_i}$  which are trivial over the systems $E$. Clearly then the stabilisers are untouched, so  the resulting state can be written

\begin{equation}
\label{eq:EC_Noisy_State}
S_{\mathcal{F}(S_{L})} = \langle f_{1} ,\dots, f_{l}, h_{L_{1}}^\star \dots h_{L_k}^\star \rangle ,
\end{equation}
with $l \leq n-k$. For now we do not say anything about how the error has acted over the code generators resulting in the $f_i$ above. This will be dealt with in the next steps, the syndrome measurements (which project the encoding onto one of several orthogonal spaces), and the correction (which take them back to the original code space),

The syndrome extraction is then performed by measuring the $n-k$ original code's generators, i.e. $g_{i} \in S_{C}$. The binary measurement outcomes are then stored in the syndrome vector $\vec{\mu}$ where

\begin{equation}
\mu_{i} \in \{ +1, -1\}, \text{ and } |\vec{\mu}| = n-k.
\end{equation}

Each $\mu_{i}$ is the outcome associated to the measurement of the code's generators $g_{i}$ on the noisy state $\mathcal{F}(\rho_{S_{L}})$. 
Since the $g_i$ commute with the $h_{L_{i}}^\star$, these remain stabilisers after the measurement. The state after the syndrome measurement is then

\begin{align}
\label{eq:Post_Syndrome_Measurement}
S_{L}^{\prime} = \langle \mu_{1}g_{1}, \dots, \mu_{n-k}g_{n-k}, h_{L_{1}}^\star \dots h_{L_k}^\star \rangle.
\end{align}

The next and final step is the correction recovery operation $R^{\vec{\mu}}$ which depends on the syndrome measurement results $\vec{\mu}$. These have two properties which make the error correction work $ R^{\vec{\mu} \dagger} \mu_i g_i R^{\vec{\mu}} = g_i$ and $ R^{\vec{\mu} \dagger} h_{L_{i}}^\star R^{\vec{\mu}} = h_{L_{i}}^\star$ for all $h_{L_{i}}^\star$. These must be true since, by assumption, this is a quantum code which has distance $d$.
One can see then that the final state is 

\begin{align}
\label{eq:Post_Syndrome_Measurement}
S_{L}^{\prime \prime} = \langle g_{1}, \dots, g_{n-k}, h_{L_{1}}^\star \dots h_{L_k}^\star \rangle,
\end{align}
which is the original encoded state. In thus way the error correction is said to have succeeded.

To see how the Toy case works one can follow through the exact same logic, with a few subtleties. 
First, one should consider a `Toy' noise transformation.
This is simple, the model of noise we consider (see section~\eqref{sec:Generalized_Transformations}) corresponds to some quantum CPTP map (of course this is not unique, but that is ok). The key point is that if the Toy noise is only non-tivial  over a set of systems $E$, the same is true for the quantum case. 

Given a toy state to encode, one considers a choice of map to the quantum case (again, any is ok).  Next, the application of Lemma 6. This is also straightfoward as the difference in the Toy and quantum cases only gives a phase, so if the product of opertors is trival (identity) in the quantum case, it is also in the Toy case.

Finally the recovery operation. Here we recall that the quantum recovery operators are Pauli. In this case the translation from quantum to Toy is straightforward (see section...), since if two quantum Pauli's commute, the corresponding toy transformation and measurement (stabiliser) operators also commute.

\section{Measurement Based Toy Computation} \label{App:MBTC}

In order to define MBTC three ingredients are needed: toy graph states, an `universal set of gates', and a deterministic correction strategy.

The universal resource state correspond to the standard cluster state, whose existence was already implicitly proven by Pusey in \cite{Pusey2012}.

An universal set of gates sufficient to ensure universal toy computation may be provided by the following

\begin{equation}
\label{eq:universal_set_of_gates}
\{ \textit{CNOT}, \textit{Hadamard}, \textit{Phase}, \tilde{\mathcal{X}}, \tilde{\mathcal{Y}}, \tilde{\mathcal{Z}}\},
\end{equation}
where $\textit{CNOT}$ is the only transformation involving two systems and consists of a control-$\tilde{\mathcal{X}}$ between a control and a target systems \footnote{In general, a control operation corresponds to the following. If the state of the control system is $0$, nothing is done to the target. On the other hand, when the state of the control is $1$, a particular transformation is applied to the target system.}, while the single system permutations toy Hadamard, $\tilde{\mathcal{H}}$, consists of a $(1)(32)(4)$ ontic permutation, the toy phase, $\tilde{\mathcal{P}}$, corresponds to a $4$-cycle $(1423)$, and $\tilde{\mathcal{X}}, \tilde{\mathcal{Y}}, \tilde{\mathcal{Z}}$ are the toy Pauli transformations and correspond to permutations $(13)(24)$, $(14)(23)$, and $(12)(34)$ respectively. All these gates can be implemented in the quantum scenario via stabilizer measurements alone and can hence straightforwardly be translated to the toy model.

Finally, we need to show that the gflow correction strategy can be implemented in the toy model. Introduced in the context of MBQC \cite{Browne}, the power of gflow lies in the idea of accounting for the randomness of quantum measurements by creating a correction strategy based on the application of the stabilizer's equations\footnote{$K_i \ket{G} = \ket{G}$, where $\ket{G}$ is the quantum graph state defined as the simultaneous eigenstate for all the generators $K_i = X_i \otimes_{j \in N(i)} Z_j$, where $j$ is in the neighbourhood of $i$ and $X$ and $Z$ are the usual pauli matrices.} which preserves the information through the computation by imposing a temporal order over the measurements. In general a MBQC computation takes into account the randomness of measurements by incorporating the corrections into the future measurement basis for yet unmeasured qubits. Gflow can be used to do this, as it provides a recipe to use the stabilizers in order to correct for the measurement's randomness. The first step to understand how gflow works in MBQC is noticing that in the $X-Y$ plane orthogonal projections are related one to another via $Z$ pauli operators, we can in fact easily show that $P_{+ \theta} =  Z P_{- \theta} Z$ where $P_{\pm \theta} := \ket{\pm_{\theta}} \bra{\pm_{\theta}}$. Now imagine we had obtained a $-1$ outcome for a measurement on qubit $i$. All we would need to do would be to apply a $Z$ operator to that qubit \textit{before} the measurement, and then we would be sure we would indeed obtain the $+1$ results. Surely enough, the law of physics do not allow us to either go back in time to correct before we measure nor to know the outcome of a measurement prior the measurement's realization but we can find a way to simulate such a strategy using the stabilizers properties. Suppose we have a graph state $\ket{G(\psi)}$ and want to apply a $Z_i$ correction on qubit $i$, then

\begin{eqnarray}
\label{eq:AnachronicalCorrection}
Z_i \ket{G(\psi)} &=& Z_i  K_{j=N(i)} \ket{G(\psi)}\\
 &=& \mathbb{I}_i \otimes X_j \otimes_{k \in N(j) \not = i} Z_k \ket{G(\psi)},
\end{eqnarray}

and since $ X_j \otimes_{k \in N(j) \not = i} Z_k $ commutes with the measurement (as they act on different qubits), then the we can apply the $\mathbb{I}\otimes X_j \otimes_{k \in N(j) \not = i} Z_k $ correction if we measured $-1$ on qubit $i$ and this would be equivalent to having performed an \textit{anachronical} $Z$ correction on qubit $i$ prior the measurement.
\medskip

Now that we have seen how gflow works in the quantum case, let us see how to translate it to the toy model. In order to define gflow in the toy model we have to prove two simple statement. \textit{a)} We need to prove that there is a \textit{projector rule} equivalent to the quantum one in the toy model. \textit{b)} We need to show that the anachronical correction also holds within the toy model framework.
\smallskip

Let us analyse the two propositions more in details:

\begin{itemize}
	\item \textit{Projector Rule}. It can be easily checked that the negative $\mathcal{X}$ and $\mathcal{Y}$ projectors are related to their positive counterparts via a similar expression to the quantum case. Letting $P_{(\pm \mathcal{X})} = 1/2 (\mathcal{I} \pm \mathcal{X})$ be the usual $\mathcal{X}$ toy projector where the $\pm$ label refers to the positive and negative projectors respectively, we can see by explicitly carrying out the matrix multiplication that $P_{(+\mathcal{X})} = \tilde{\mathcal{Z}} P_{(-\mathcal{X})} \tilde{\mathcal{Z}}$. Similarly for the $\mathcal{Y}$ projector, as $P_{(+\mathcal{Y})} = \tilde{\mathcal{Z}} P_{(- \mathcal{Y})} \tilde{\mathcal{Z}}$. This means that if we consider the $\mathcal{X}$ and $\mathcal{Y}$ projectors the equivalent of the $X-Y$ plane measurements for the toy model, we can related the negative and the positive outcomes of measurements one to another by  means of a $\tilde{\mathcal{Z}}$ permutation on the system of interest.
	\item \textit{Anachronical Correction}. In order to apply the $\tilde{\mathcal{X}}_j \otimes_{k \in N(j) \not = i} \tilde{\mathcal{Z}}_k $ correction, we need to show that the equation~\eqref{eq:AnachronicalCorrection} holds for both toy observables and toy permutations. That is, we want to show that it works for both $\mathcal{Z}_i$ and $\tilde{\mathcal{Z}}_{i}$. Equation~\eqref{eq:AnachronicalCorrection} will straightforwardly hold in both the toy-stabilizers and the toy-permutation case provided that $\mathcal{K}_i \mathcal{K}_i = \mathbb{I}$  and $\tilde{\mathcal{K}}_i \tilde{\mathcal{K}}_i = \mathbb{I}$. And these two conditions are trivially true by definition. Moreover, it is important to note that since operations on different systems commute, we can freely apply the $\tilde{\mathcal{X}}_j \otimes_{k \in N(j) \not = i} \tilde{\mathcal{Z}}_k $ correction either before or after the measurement $\mathcal{M}_i$.

\end{itemize}

We can then provide a formal definition of gflow reminiscent of the one in \cite{Browne}

\begin{definition}
\textit{An open toy graph state $G(I,O,V)$ with $n$ vertexes and input set $I$ and output set $O$ has gflow if there exists a map $g: O^{C} \rightarrow P^{I^{C}}$ (from measured toy systems to a subset of the prepared toy systems) and a partial order $\prec$ over $V$ for all $i \in O^{C}$ such that 
\begin{itemize}
	\item[g(1)] if $j \in g(i)$ and $i \not = j$ then $i \prec j$,
	\item[g(2)] - if $j \prec i$ and $i \not = j$ then $j \not \in Odd(g(i))$,
	\item[g(3)] $i \not \in g(i)$ and $i \in Odd(g(i))$.
\end{itemize}
}
\end{definition}

These conditions assure the consistency of the correction. In particular, $g(1)$ ensures  each correction happening after the measurement and not before it; $g(2)$  ensures the gflow of the $i$-th system to be evenly connected to the measured systems (i.e. the past) so that there is never a single $\tilde{\mathcal{Z}}$ acting upon the past of the computation hence changing the state of the already measured systems; finally, $g(3)$ forces the gflow to be oddly connected to the $i$-th system as this assures that the correction strategy leaves the required $\tilde{\mathcal{Z}}$ correction on the $i$-th system. Hence we conclude that for any quantum graph state which has gflow, we can construct an equivalent toy graph state with toy gflow. Thus, the toy gflow provides an implementable correcting scheme to assure deterministic measurement based toy computation(MBTC).

\section{Proof of Theorem \ref{thm:Blindness} (Blindness)} \label{app:Blindness_proof}

The correctness of the protocol is easily recovered as the two only differences between blind and non-blind delegated MBTC protocols are the random initial $\theta$ masking and the `$r$' one-time padded instructions that Alice sends to Bob. Now, the role of the $\delta  = (\phi_{1} \oplus \theta_{1} \oplus r) + 2(\theta_{2} \oplus  \phi_{2}) $ instruction is exactly that of removing the initial mask while hiding the true measurement setting hence performing the correct measurement. Furthermore, the addition of the $r$ only flips the meaning of the outcomes: if the outcome of the $i$-th measurement is $o_{i}$ all Alice has to do in order to decode the real measurement outcome is to calculate $o_{i}^{\prime} = o_{i} \oplus r_{i}$. On the other hand, blindness can be proved by showing that no information (beside the size of the whole computation) is leaked at any stage to Bob: not from the used resource state, not during the computation itself, nor from the output state. First, the resource state leaks no information as the use of a universal toy graph leaks no information about the computation besides an upper bound on the total number of systems. In order to show that Bob is blind on the final output state and at any stage of the computation considering the $i$-th system.
In the protocol Alice masks the real measurement setting by asking Bob to measure with setting $\delta  = (\phi_{1} \oplus \theta_{1} \oplus r) + 2(\theta_{2} \oplus  \phi_{2})$
Since $\theta_{i}$ and $r_i$ are chosen uniformly at random we can see that the randomness of $\theta_{i}$ decorrelates entirely the real measurement setting $\phi_{i}'$ from the one communicated to Bob by $\delta_{i}$. On the other hand, summing over $r_{i}$ implies that the measurement outcomes corresponding to the masked angle $\phi'_{i} + \theta_{i}$ are also completely randomized as

\begin{equation}
\sum_{r_{i}} p(r_{i}) Tr(  P_{(\phi'_{i} + \theta_{i})} \tilde{\mathcal{Z}}^{r_{i}} \rho_{(\phi'_{i} + \theta_{i} )} \tilde{\mathcal{Z}}^{r_{i}}) = 1/2,
\end{equation}
where $ p(r_{i}) = 1/2$ represents the probabilities associated to the two values of $r_{i}$. \qed

\section{Proof of Theorem \ref{thm:Verification_Theorem} (Verification)} \label{app:Verification_proof}

Let equation \eqref{eq:Verification_Prob_Failiure} be divided into two parts

\begin{multline}
\label{eq:Verification_Bound_A_and_B}
p_{fail} = \sum_{t,r_{t},\nu_{c}} p(t,r_{t},\nu_{c}) [ \overbrace{Tr_{O \setminus t, t}( I_{O \setminus t} \otimes P^{r_{t}}_{t}) \rho^{out}_{\Ot,t} )}^{\alpha} - \\ 
\underbrace{Tr_{O \setminus t, t} (P^{honest}_{\Ot,t} (\rho^{out}_{O \setminus t,t})}_{\beta} ].
\end{multline}

Intuitively this breaks down the probability of failure into the probability of springing a trap (\textit{alpha}), minus the probability of having corrupted the computation (\textit{beta}). Note further that effectively $P^{honest}$ in~\eqref{eq:Verification_Bound_A_and_B} is defined as $P^{honest}_{\Ot,t} = P^{honest}_{\Ot,t} \otimes P^{r_{t}}_{t}$, that is over outputs and the traps. As both $\alpha$ and $\beta$ are probabilities, a sufficient and necessary condition to obtain $p_{fail} = 1$ is to have $\sum_{t,r_{t},\nu_{c}} p(t,r_{t},\nu_{c}) \alpha = 1$ and simultaneously $\sum_{t,r_{t},\nu_{c}} p(t,r_{t},\nu_{c}) \beta = 0$. That is, never springing a trap but also corrupting the computation.

Consider first the condition on $\alpha$. By taking the expression of $\rho^{out}_{O \setminus t,t}$ given by equation~\eqref{eq:Verification_Output_Computation} (note the partial traces over $\Delta$ and $B$ are already been performed) the condition reads

\begin{align}
\alpha &= \sum_{\vec{b}}  p(\vec{b}) Tr_{N} \big[ (I_{O \setminus t} \otimes P^{r_{t}}_{t}) [ \rho^{\vec{b}}_{\N} \otimes \tilde{\mathcal{C}}_{\Ot}^{\bNu} (\sigma_{{\Ot,t}}^{\prime \vec{b}}) \tilde{\mathcal{C}}_{\Ot}^{\bNu} ] \big] \nonumber \\
&= 1. \nonumber
\end{align}

Then using the cyclic properties of the trace the corrections $\tilde{\mathcal{C}}_{\Ot}^{\bNu}$ cancel, and the partial trace over $\N$ can then be taken, yielding

\begin{equation}
\alpha = \sum_{\vec{b}}  p(\vec{b}) Tr_{\Ot, t} \big[(I_{O \setminus t} \otimes P^{r_{t}}_{t}) (\sigma_{{\Ot,t}}^{\prime \vec{b}}) \big].
\end{equation}

Then, by letting $\sigma_{{t}}^{\prime \prime \vec{b}} = Tr_{\Ot} (\sigma_{{\Ot,t}}^{\prime \vec{b}})$ a \textit{necessary} condition to achieve $\alpha = 1$ is given by

\begin{equation}
\label{eq:Verification_Condition_From_Alpha1}
Tr_{t} (P^{r_{t}}_{t} \sigma_{{t}}^{\prime \prime \vec{b}}) = 1.
\end{equation}

This condition needs to be valid for all trap's position $t \in N$, all $r_{i} \in \{0,1\}$, and, crucially, for all branches of the computation $\vec{b}$. That is, for all those parameters $\sigma_{{t}}^{\prime \prime \vec{b}}$ has to be the correct trap state as Alice initialized it. This poses a strong condition on the allowed form of Bob's deviation $\tilde \Omega^{All}$. Note the state $\sigma_{{\Ot,t}}^{\prime \vec{b}}$ (and subsequently $\sigma_{{t}}^{\prime \prime \vec{b}}$) is obtained only via the $\N$ projective measurements and tracing out operations. Recalling the updates rules given in~\eqref{sec:Measurements} one can thus observe that the stabilizer of the resulting state $S_{\sigma^{\prime}}$ is generated by considering $\N$ generators encoding the measurements along with up to $\Ot,t$ linearly independent generators from the $S_{\sigma^{All}}$ (obtained from equation~\eqref{eq:First_Output_state_in_full}). However both the state-update rule and the partial trace only add non-trivial stabilizers on the $\N$ systems and not on the trap. That is, if equation~\eqref{eq:Verification_Condition_From_Alpha1} is to hold for all $t,r_{t},$ and $\vec{b}$ it must be the case that
 
\begin{align}
\label{eq:Verification_Condition_From_Alpha}
\sigma_{All} &= \bar \sigma_{N \setminus t,\Delta \setminus t,B} \otimes \rho^{r_{t}}_{t} \otimes \delta_{t}
\end{align}
where $ \bar \sigma_{\N,\Delta \setminus t ,B}$ is an arbitrary state over all non traps systems, and $\rho^{r_{t}}_{t}$ is the state of the trap as Alice initialized it and $\delta_{t}$ is the corresponding measurement instruction. In other words equation~\eqref{eq:Verification_Condition_From_Alpha} explicitly constrains Bob's deviation insofar as it requires to preserve the tensor product structure between the traps and all other systems.

Now consider the condition on $\beta$. Tracing out $\Delta$ and $B$ the expression reads

\begin{align}
\beta &= \sum_{\vec{b}}  p(\vec{b}) Tr_{(N)} \big[ P^{honest}_{\Ot,t}[ \rho^{\vec{b}}_{\N} \otimes \tilde{C}_{\Ot}^{\bNu}  (\sigma_{{\Ot,t}}^{\prime \vec{b}}) \tilde{C}_{\Ot}^{\bNu} ] \big]  \nonumber \\ 
&= 0.
\end{align}

As for the alpha terms a condition on beta is obtained performing the partial trace over the $\N $ systems hence obtaining

\begin{align}
\beta &= \sum_{\vec{b}}  p(\vec{b}) Tr_{\Ot,t} \big[ P^{honest}_{\Ot,t} [ \tilde{\mathcal{C}}_{\Ot}^{\bNu} \sigma_{ {\Ot,t}}^{\prime \vec{b}} \tilde{\mathcal{C}}_{\Ot}^{\bNu} ] \big] \nonumber \\
 &= 0. 
\end{align}

This also identifies a necessary condition on the state $\tilde{\mathcal{C}}_{\Ot}^{\bNu} \sigma_{{\Ot,t}}^{\prime \vec{b}} \tilde{\mathcal{C}}_{\Ot}^{\bNu}$. Namely it must be the case that

\begin{equation}
\label{eq:Verification_Condition_From_Beta}
\forall \; \vec{b}, \; \exists \; \mathcal Z_{i} \in S_{honest} 
\; s.t. \;  - \mathcal Z_{i} \in S_{\sigma^{\prime}}, \text{ for } i \in (\Ot,t),
\end{equation}
where $S_{\sigma^{\prime \prime}}$ is the stabilizer group of the state $\tilde{\mathcal{C}}_{\Ot}^{\bNu} \sigma_{{\Ot,t}}^{\prime \prime \vec{b}} \tilde{\mathcal{C}}_{\Ot}^{\bNu}$, and $S^{honest}$ is defined as in equation~\eqref{eq:Verification_Honest_Computation} over both the outputs and the trap systems. However condition~\eqref{eq:Verification_Condition_From_Beta} is not consistent with~\eqref{eq:Verification_Condition_From_Alpha} because it implies the existence of a position where the traps state is changed from $r_{r}$ to $r_{t} \oplus 1$.


In order to find a bound we can relax Bob's strategy to see how close he can get to $p_{fail}=1$. This can be calculated by taking the condition $\beta = 0$ to hold, by having a cheating strategy which produces a change in only one of the generators of $S_{\sigma^{\prime}}$. At position $T$ then let the deviation result in either $\pm \mathcal{X}_{T}$ or $\pm \mathcal{Y}_{T}$. Then $ \sum_{r_{t}} Tr_{t=T} (P_{r_{t}} \rho_{\mathcal{X}}) = 1$ when the sum over $r_{t}$ is performed. Therefore when as the sum is taken over all the random parameters $p_{fail} = (1/2n)(2n - 1) = 1 - 1/2n$. \qed

\bibliography{ToyPaper}

\begin{thebibliography}{37}%
\makeatletter
\providecommand \@ifxundefined [1]{%
 \@ifx{#1\undefined}
}%
\providecommand \@ifnum [1]{%
 \ifnum #1\expandafter \@firstoftwo
 \else \expandafter \@secondoftwo
 \fi
}%
\providecommand \@ifx [1]{%
 \ifx #1\expandafter \@firstoftwo
 \else \expandafter \@secondoftwo
 \fi
}%
\providecommand \natexlab [1]{#1}%
\providecommand \enquote  [1]{``#1''}%
\providecommand \bibnamefont  [1]{#1}%
\providecommand \bibfnamefont [1]{#1}%
\providecommand \citenamefont [1]{#1}%
\providecommand \href@noop [0]{\@secondoftwo}%
\providecommand \href [0]{\begingroup \@sanitize@url \@href}%
\providecommand \@href[1]{\@@startlink{#1}\@@href}%
\providecommand \@@href[1]{\endgroup#1\@@endlink}%
\providecommand \@sanitize@url [0]{\catcode `\\12\catcode `\$12\catcode
  `\&12\catcode `\#12\catcode `\^12\catcode `\_12\catcode `\%12\relax}%
\providecommand \@@startlink[1]{}%
\providecommand \@@endlink[0]{}%
\providecommand \url  [0]{\begingroup\@sanitize@url \@url }%
\providecommand \@url [1]{\endgroup\@href {#1}{\urlprefix }}%
\providecommand \urlprefix  [0]{URL }%
\providecommand \Eprint [0]{\href }%
\providecommand \doibase [0]{http://dx.doi.org/}%
\providecommand \selectlanguage [0]{\@gobble}%
\providecommand \bibinfo  [0]{\@secondoftwo}%
\providecommand \bibfield  [0]{\@secondoftwo}%
\providecommand \translation [1]{[#1]}%
\providecommand \BibitemOpen [0]{}%
\providecommand \bibitemStop [0]{}%
\providecommand \bibitemNoStop [0]{.\EOS\space}%
\providecommand \EOS [0]{\spacefactor3000\relax}%
\providecommand \BibitemShut  [1]{\csname bibitem#1\endcsname}%
\let\auto@bib@innerbib\@empty
\bibitem [{\citenamefont {Spekkens}(2007)}]{Spekkens2007}%
  \BibitemOpen
  \bibfield  {author} {\bibinfo {author} {\bibfnamefont {R.~W.}\ \bibnamefont
  {Spekkens}},\ }\href {\doibase 10.1103/PhysRevA.75.032110} {\bibfield
  {journal} {\bibinfo  {journal} {Physical Review A}\ }\textbf {\bibinfo
  {volume} {75}},\ \bibinfo {pages} {032110} (\bibinfo {year}
  {2007})}\BibitemShut {NoStop}%
\bibitem [{\citenamefont {Spekkens}(2016)}]{Spekkens2016}%
  \BibitemOpen
  \bibfield  {author} {\bibinfo {author} {\bibfnamefont {R.~W.}\ \bibnamefont
  {Spekkens}},\ }\href@noop {} {\emph {\bibinfo {title} {Quantum Theory:
  Informational Foundations and Foils}}}\ (\bibinfo  {publisher} {Springer
  Netherlands},\ \bibinfo {year} {2016})\ pp.\ \bibinfo {pages}
  {83--135}\BibitemShut {NoStop}%
\bibitem [{\citenamefont {Coecke}\ \emph {et~al.}(2011)\citenamefont {Coecke},
  \citenamefont {Edwards},\ and\ \citenamefont {Spekkens}}]{Coecke2011}%
  \BibitemOpen
  \bibfield  {author} {\bibinfo {author} {\bibfnamefont {B.}~\bibnamefont
  {Coecke}}, \bibinfo {author} {\bibfnamefont {B.}~\bibnamefont {Edwards}}, \
  and\ \bibinfo {author} {\bibfnamefont {R.~W.}\ \bibnamefont {Spekkens}},\
  }\href@noop {} {\bibfield  {journal} {\bibinfo  {journal} {Electronic Notes
  in Theoretical Computer Science}\ }\textbf {\bibinfo {volume} {270}},\
  \bibinfo {pages} {15} (\bibinfo {year} {2011})}\BibitemShut {NoStop}%
\bibitem [{\citenamefont {Pusey}(2012)}]{Pusey2012}%
  \BibitemOpen
  \bibfield  {author} {\bibinfo {author} {\bibfnamefont {M.~F.}\ \bibnamefont
  {Pusey}},\ }\href {\doibase 10.1007/s10701-012-9639-7} {\bibfield  {journal}
  {\bibinfo  {journal} {Foundations of Physics}\ }\textbf {\bibinfo {volume}
  {42}},\ \bibinfo {pages} {688} (\bibinfo {year} {2012})}\BibitemShut
  {NoStop}%
\bibitem [{\citenamefont {Raussendorf}\ and\ \citenamefont
  {Briegel}(2001)}]{Raussendorf2001}%
  \BibitemOpen
  \bibfield  {author} {\bibinfo {author} {\bibfnamefont {R.}~\bibnamefont
  {Raussendorf}}\ and\ \bibinfo {author} {\bibfnamefont {H.}~\bibnamefont
  {Briegel}},\ }\href {\doibase http://dx.doi.org/10.1103/PhysRevLett.86.5188}
  {\bibfield  {journal} {\bibinfo  {journal} {Physical Review Letters}\
  }\textbf {\bibinfo {volume} {86}},\ \bibinfo {pages} {5188} (\bibinfo {year}
  {2001})}\BibitemShut {NoStop}%
\bibitem [{\citenamefont {Gottesman}(1997)}]{Gottesman1997}%
  \BibitemOpen
  \bibfield  {author} {\bibinfo {author} {\bibfnamefont {D.}~\bibnamefont
  {Gottesman}},\ }\href {http://arxiv.org/abs/quant-ph/9705052} {\bibfield
  {journal} {\bibinfo  {journal} {Caltech Ph.D. Thesis}\ } (\bibinfo {year}
  {1997})},\ \Eprint {http://arxiv.org/abs/9705052} {arXiv:9705052}
  \BibitemShut {NoStop}%
\bibitem [{\citenamefont {Broadbent}\ \emph {et~al.}(2008)\citenamefont
  {Broadbent}, \citenamefont {Fitzsimons},\ and\ \citenamefont
  {Kashefi}}]{Broadbent2008}%
  \BibitemOpen
  \bibfield  {author} {\bibinfo {author} {\bibfnamefont {A.}~\bibnamefont
  {Broadbent}}, \bibinfo {author} {\bibfnamefont {J.}~\bibnamefont
  {Fitzsimons}}, \ and\ \bibinfo {author} {\bibfnamefont {E.}~\bibnamefont
  {Kashefi}},\ }in\ \href {http://arxiv.org/abs/0807.4154} {\emph {\bibinfo
  {booktitle} {Foundations of Computer Science, 2009. FOCS'09. 50th Annual IEEE
  Symposium on}}}\ (\bibinfo {year} {2008})\ pp.\ \bibinfo {pages}
  {517--526}\BibitemShut {NoStop}%
\bibitem [{\citenamefont {Fitzsimons}\ and\ \citenamefont
  {Kashefi}(2012)}]{Fitzsimons2012}%
  \BibitemOpen
  \bibfield  {author} {\bibinfo {author} {\bibfnamefont {J.~F.}\ \bibnamefont
  {Fitzsimons}}\ and\ \bibinfo {author} {\bibfnamefont {E.}~\bibnamefont
  {Kashefi}},\ }\href@noop {} {\bibfield  {journal} {\bibinfo  {journal} {arXiv
  preprint arXiv:1203.5217}\ } (\bibinfo {year} {2012})}\BibitemShut {NoStop}%
\bibitem [{\citenamefont {Kochen}\ and\ \citenamefont
  {Specker}(1967)}]{Kochen1967}%
  \BibitemOpen
  \bibfield  {author} {\bibinfo {author} {\bibfnamefont {S.}~\bibnamefont
  {Kochen}}\ and\ \bibinfo {author} {\bibfnamefont {E.~P.}\ \bibnamefont
  {Specker}},\ }\href@noop {} {\bibfield  {journal} {\bibinfo  {journal}
  {Journal of Mathematics and Mechanics}\ }\textbf {\bibinfo {volume} {17}},\
  \bibinfo {pages} {59} (\bibinfo {year} {1967})}\BibitemShut {NoStop}%
\bibitem [{\citenamefont {Bartlett}\ \emph {et~al.}(2012)\citenamefont
  {Bartlett}, \citenamefont {Rudolph},\ and\ \citenamefont
  {Spekkens}}]{Bartlett2012}%
  \BibitemOpen
  \bibfield  {author} {\bibinfo {author} {\bibfnamefont {S.~D.}\ \bibnamefont
  {Bartlett}}, \bibinfo {author} {\bibfnamefont {T.}~\bibnamefont {Rudolph}}, \
  and\ \bibinfo {author} {\bibfnamefont {R.~W.}\ \bibnamefont {Spekkens}},\
  }\href@noop {} {\bibfield  {journal} {\bibinfo  {journal} {Physical Review
  A}\ }\textbf {\bibinfo {volume} {86}},\ \bibinfo {pages} {012103} (\bibinfo
  {year} {2012})}\BibitemShut {NoStop}%
\bibitem [{\citenamefont {Weedbrook}\ \emph {et~al.}(2012)\citenamefont
  {Weedbrook}, \citenamefont {Pirandola}, \citenamefont {Garcia-Patron},
  \citenamefont {Cerf}, \citenamefont {Ralph}, \citenamefont {Shapiro},\ and\
  \citenamefont {Lloyd}}]{GaussianQuantumInformation}%
  \BibitemOpen
  \bibfield  {author} {\bibinfo {author} {\bibfnamefont {C.}~\bibnamefont
  {Weedbrook}}, \bibinfo {author} {\bibfnamefont {S.}~\bibnamefont
  {Pirandola}}, \bibinfo {author} {\bibfnamefont {R.}~\bibnamefont
  {Garcia-Patron}}, \bibinfo {author} {\bibfnamefont {N.~J.}\ \bibnamefont
  {Cerf}}, \bibinfo {author} {\bibfnamefont {T.~C.}\ \bibnamefont {Ralph}},
  \bibinfo {author} {\bibfnamefont {J.~H.}\ \bibnamefont {Shapiro}}, \ and\
  \bibinfo {author} {\bibfnamefont {S.}~\bibnamefont {Lloyd}},\ }\href
  {\doibase 10.1103/RevModPhys.84.621} {\bibfield  {journal} {\bibinfo
  {journal} {Rev. Mod. Phys.}\ }\textbf {\bibinfo {volume} {84}},\ \bibinfo
  {pages} {621} (\bibinfo {year} {2012})}\BibitemShut {NoStop}%
\bibitem [{\citenamefont {Magnin}\ \emph {et~al.}(2010)\citenamefont {Magnin},
  \citenamefont {Magniez}, \citenamefont {Leverrier},\ and\ \citenamefont
  {Cerf}}]{Leverrier2014}%
  \BibitemOpen
  \bibfield  {author} {\bibinfo {author} {\bibfnamefont {L.}~\bibnamefont
  {Magnin}}, \bibinfo {author} {\bibfnamefont {F.}~\bibnamefont {Magniez}},
  \bibinfo {author} {\bibfnamefont {A.}~\bibnamefont {Leverrier}}, \ and\
  \bibinfo {author} {\bibfnamefont {N.~J.}\ \bibnamefont {Cerf}},\ }\href
  {\doibase 10.1103/PhysRevA.81.010302} {\bibfield  {journal} {\bibinfo
  {journal} {Phys. Rev. A}\ }\textbf {\bibinfo {volume} {81}},\ \bibinfo
  {pages} {010302} (\bibinfo {year} {2010})}\BibitemShut {NoStop}%
\bibitem [{\citenamefont {Backens}\ and\ \citenamefont
  {Duman}(2015)}]{Backens2015}%
  \BibitemOpen
  \bibfield  {author} {\bibinfo {author} {\bibfnamefont {M.}~\bibnamefont
  {Backens}}\ and\ \bibinfo {author} {\bibfnamefont {A.~N.}\ \bibnamefont
  {Duman}},\ }\href {\doibase 10.1007/s10701-015-9957-7} {\bibfield  {journal}
  {\bibinfo  {journal} {Foundations of Physics}\ }\textbf {\bibinfo {volume}
  {46}},\ \bibinfo {pages} {70} (\bibinfo {year} {2015})}\BibitemShut {NoStop}%
\bibitem [{\citenamefont {{Nielsen, Michael A. and
  Chuang}}(2010)}]{NielsenMichaelA.andChuang2010}%
  \BibitemOpen
  \bibfield  {author} {\bibinfo {author} {\bibfnamefont {I.~L.}\ \bibnamefont
  {{Nielsen, Michael A. and Chuang}}},\ }\href@noop {} {\emph {\bibinfo {title}
  {{Quantum computation and quantum information}}}},\ edited by\ \bibinfo
  {editor} {\bibfnamefont {C.~U.}\ \bibnamefont {Press;}}\ (\bibinfo {year}
  {2010})\BibitemShut {NoStop}%
\bibitem [{Note1()}]{Note1}%
  \BibitemOpen
  \bibinfo {note} {The Clifford group is the group of unitary operations that
  maps the Pauli group to itself under conjugation.}\BibitemShut {Stop}%
\bibitem [{Note2()}]{Note2}%
  \BibitemOpen
  \bibinfo {note} {E.g. the notation $(13)(24)$ means that $1 \rightarrow 3$,
  $3 \rightarrow 1 $ and $2 \rightarrow 4$, $4 \rightarrow 2$.}\BibitemShut
  {Stop}%
\bibitem [{\citenamefont {Catani}\ and\ \citenamefont
  {Browne}(2017)}]{Catani2017}%
  \BibitemOpen
  \bibfield  {author} {\bibinfo {author} {\bibfnamefont {L.}~\bibnamefont
  {Catani}}\ and\ \bibinfo {author} {\bibfnamefont {D.~E.}\ \bibnamefont
  {Browne}},\ }\href@noop {} {\bibfield  {journal} {\bibinfo  {journal} {arXiv
  preprint arXiv:1701.07801}\ } (\bibinfo {year} {2017})}\BibitemShut {NoStop}%
\bibitem [{\citenamefont {Pati}\ and\ \citenamefont
  {Braunstein}(2000)}]{Pati2000}%
  \BibitemOpen
  \bibfield  {author} {\bibinfo {author} {\bibfnamefont {A.~K.}\ \bibnamefont
  {Pati}}\ and\ \bibinfo {author} {\bibfnamefont {S.~L.}\ \bibnamefont
  {Braunstein}},\ }\href@noop {} {\bibfield  {journal} {\bibinfo  {journal}
  {Nature}\ }\textbf {\bibinfo {volume} {404}},\ \bibinfo {pages} {164}
  (\bibinfo {year} {2000})}\BibitemShut {NoStop}%
\bibitem [{\citenamefont {Lo}\ and\ \citenamefont {Chau}(1997)}]{Lo1997}%
  \BibitemOpen
  \bibfield  {author} {\bibinfo {author} {\bibfnamefont {H.-K.}\ \bibnamefont
  {Lo}}\ and\ \bibinfo {author} {\bibfnamefont {H.~F.}\ \bibnamefont {Chau}},\
  }\href {http://arxiv.org/abs/quant-ph/9603004} {\bibfield  {journal}
  {\bibinfo  {journal} {Physical Review Letters}\ }\textbf {\bibinfo {volume}
  {78}},\ \bibinfo {pages} {3410} (\bibinfo {year} {1997})}\BibitemShut
  {NoStop}%
\bibitem [{\citenamefont {Mayers}(1997)}]{Mayers1997}%
  \BibitemOpen
  \bibfield  {author} {\bibinfo {author} {\bibfnamefont {D.}~\bibnamefont
  {Mayers}},\ }\href {\doibase http://dx.doi.org/10.1103/PhysRevLett.78.3414}
  {\bibfield  {journal} {\bibinfo  {journal} {Physical Review Letters}\
  }\textbf {\bibinfo {volume} {78}},\ \bibinfo {pages} {3414} (\bibinfo {year}
  {1997})}\BibitemShut {NoStop}%
\bibitem [{\citenamefont {D'Ariano}\ \emph {et~al.}(2007)\citenamefont
  {D'Ariano}, \citenamefont {Kretschmann}, \citenamefont {Schlingemann},\ and\
  \citenamefont {Werner}}]{Dariano2007}%
  \BibitemOpen
  \bibfield  {author} {\bibinfo {author} {\bibfnamefont {G.~M.}\ \bibnamefont
  {D'Ariano}}, \bibinfo {author} {\bibfnamefont {D.}~\bibnamefont
  {Kretschmann}}, \bibinfo {author} {\bibfnamefont {D.}~\bibnamefont
  {Schlingemann}}, \ and\ \bibinfo {author} {\bibfnamefont {R.~F.}\
  \bibnamefont {Werner}},\ }\href {\doibase 10.1103/PhysRevA.76.032328}
  {\bibfield  {journal} {\bibinfo  {journal} {Phys. Rev. A}\ }\textbf {\bibinfo
  {volume} {76}},\ \bibinfo {pages} {032328} (\bibinfo {year}
  {2007})}\BibitemShut {NoStop}%
\bibitem [{\citenamefont {Cleve}\ \emph {et~al.}(1999)\citenamefont {Cleve},
  \citenamefont {Gottesman},\ and\ \citenamefont {Lo}}]{Cleve1999}%
  \BibitemOpen
  \bibfield  {author} {\bibinfo {author} {\bibfnamefont {R.}~\bibnamefont
  {Cleve}}, \bibinfo {author} {\bibfnamefont {D.}~\bibnamefont {Gottesman}}, \
  and\ \bibinfo {author} {\bibfnamefont {H.-K.}\ \bibnamefont {Lo}},\
  }\href@noop {} {\bibfield  {journal} {\bibinfo  {journal} {Physical Review
  Letters}\ }\textbf {\bibinfo {volume} {83}},\ \bibinfo {pages} {648}
  (\bibinfo {year} {1999})}\BibitemShut {NoStop}%
\bibitem [{\citenamefont {Markham}\ and\ \citenamefont
  {Marin}(2015)}]{Marin2014}%
  \BibitemOpen
  \bibfield  {author} {\bibinfo {author} {\bibfnamefont {D.}~\bibnamefont
  {Markham}}\ and\ \bibinfo {author} {\bibfnamefont {A.}~\bibnamefont
  {Marin}},\ }in\ \href@noop {} {\emph {\bibinfo {booktitle} {International
  Conference on Information Theoretic Security}}}\ (\bibinfo {organization}
  {Springer},\ \bibinfo {year} {2015})\ pp.\ \bibinfo {pages}
  {1--14}\BibitemShut {NoStop}%
\bibitem [{\citenamefont {Chiribella}\ \emph {et~al.}(2011)\citenamefont
  {Chiribella}, \citenamefont {D'Ariano},\ and\ \citenamefont
  {Perinotti}}]{Chiribella2011}%
  \BibitemOpen
  \bibfield  {author} {\bibinfo {author} {\bibfnamefont {G.}~\bibnamefont
  {Chiribella}}, \bibinfo {author} {\bibfnamefont {G.~M.}\ \bibnamefont
  {D'Ariano}}, \ and\ \bibinfo {author} {\bibfnamefont {P.}~\bibnamefont
  {Perinotti}},\ }\href {\doibase 10.1103/PhysRevA.84.012311} {\bibfield
  {journal} {\bibinfo  {journal} {Phys. Rev. A}\ }\textbf {\bibinfo {volume}
  {84}},\ \bibinfo {pages} {012311} (\bibinfo {year} {2011})}\BibitemShut
  {NoStop}%
\bibitem [{\citenamefont {Raussendorf}(2013)}]{Raussendorf2013}%
  \BibitemOpen
  \bibfield  {author} {\bibinfo {author} {\bibfnamefont {R.}~\bibnamefont
  {Raussendorf}},\ }\href {\doibase 10.1103/PhysRevA.88.022322} {\bibfield
  {journal} {\bibinfo  {journal} {Phys. Rev. A}\ }\textbf {\bibinfo {volume}
  {88}},\ \bibinfo {pages} {022322} (\bibinfo {year} {2013})}\BibitemShut
  {NoStop}%
\bibitem [{\citenamefont {Danos}\ \emph {et~al.}(2007)\citenamefont {Danos},
  \citenamefont {Kashefi},\ and\ \citenamefont {Panangaden}}]{Danos2007}%
  \BibitemOpen
  \bibfield  {author} {\bibinfo {author} {\bibfnamefont {V.}~\bibnamefont
  {Danos}}, \bibinfo {author} {\bibfnamefont {E.}~\bibnamefont {Kashefi}}, \
  and\ \bibinfo {author} {\bibfnamefont {P.}~\bibnamefont {Panangaden}},\
  }\href@noop {} {\bibfield  {journal} {\bibinfo  {journal} {Journal of the ACM
  (JACM)}\ }\textbf {\bibinfo {volume} {54}},\ \bibinfo {pages} {8} (\bibinfo
  {year} {2007})}\BibitemShut {NoStop}%
\bibitem [{\citenamefont {Mantri}\ \emph {et~al.}(2016)\citenamefont {Mantri},
  \citenamefont {Demarie},\ and\ \citenamefont {Fitzsimons}}]{Mantri2016}%
  \BibitemOpen
  \bibfield  {author} {\bibinfo {author} {\bibfnamefont {A.}~\bibnamefont
  {Mantri}}, \bibinfo {author} {\bibfnamefont {T.~F.}\ \bibnamefont {Demarie}},
  \ and\ \bibinfo {author} {\bibfnamefont {J.~F.}\ \bibnamefont {Fitzsimons}},\
  }\href@noop {} {\bibfield  {journal} {\bibinfo  {journal} {arXiv preprint
  arXiv:1607.00758}\ } (\bibinfo {year} {2016})}\BibitemShut {NoStop}%
\bibitem [{\citenamefont {Browne}\ \emph {et~al.}(2007)\citenamefont {Browne},
  \citenamefont {Kashefi}, \citenamefont {Mhalla},\ and\ \citenamefont
  {Perdrix}}]{Browne}%
  \BibitemOpen
  \bibfield  {author} {\bibinfo {author} {\bibfnamefont {D.~E.}\ \bibnamefont
  {Browne}}, \bibinfo {author} {\bibfnamefont {E.}~\bibnamefont {Kashefi}},
  \bibinfo {author} {\bibfnamefont {M.}~\bibnamefont {Mhalla}}, \ and\ \bibinfo
  {author} {\bibfnamefont {S.}~\bibnamefont {Perdrix}},\ }\href {\doibase
  10.1088/1367-2630/9/8/250} {\bibfield  {journal} {\bibinfo  {journal} {New
  Journal of Physics}\ }\textbf {\bibinfo {volume} {9}},\ \bibinfo {pages}
  {250} (\bibinfo {year} {2007})}\BibitemShut {NoStop}%
\bibitem [{\citenamefont {Markham}\ and\ \citenamefont
  {Kashefi}(2014)}]{Markham2014}%
  \BibitemOpen
  \bibfield  {author} {\bibinfo {author} {\bibfnamefont {D.}~\bibnamefont
  {Markham}}\ and\ \bibinfo {author} {\bibfnamefont {E.}~\bibnamefont
  {Kashefi}},\ }in\ \href@noop {} {\emph {\bibinfo {booktitle} {Horizons of the
  Mind}}}\ (\bibinfo {year} {2014})\ pp.\ \bibinfo {pages}
  {427--453}\BibitemShut {NoStop}%
\bibitem [{\citenamefont {Aharonov}\ and\ \citenamefont
  {Vazirani}(2012)}]{Aharonov2012}%
  \BibitemOpen
  \bibfield  {author} {\bibinfo {author} {\bibfnamefont {D.}~\bibnamefont
  {Aharonov}}\ and\ \bibinfo {author} {\bibfnamefont {U.}~\bibnamefont
  {Vazirani}},\ }\href@noop {} {\emph {\bibinfo {title} {Philosophy of Science
  anthology Computability: Godel, Turing, Church, and beyond}}},\ edited by\
  \bibinfo {editor} {\bibfnamefont {C.~P. S.~M.}\ \bibnamefont {press}}\
  (\bibinfo  {publisher} {June},\ \bibinfo {year} {2012})\BibitemShut {NoStop}%
\bibitem [{\citenamefont {Reichardt}\ \emph {et~al.}(2013)\citenamefont
  {Reichardt}, \citenamefont {Unger},\ and\ \citenamefont
  {Vazirani}}]{Reichardt2013}%
  \BibitemOpen
  \bibfield  {author} {\bibinfo {author} {\bibfnamefont {B.~W.}\ \bibnamefont
  {Reichardt}}, \bibinfo {author} {\bibfnamefont {F.}~\bibnamefont {Unger}}, \
  and\ \bibinfo {author} {\bibfnamefont {U.}~\bibnamefont {Vazirani}},\
  }\href@noop {} {\bibfield  {journal} {\bibinfo  {journal} {Nature}\ }\textbf
  {\bibinfo {volume} {496}},\ \bibinfo {pages} {456} (\bibinfo {year}
  {2013})}\BibitemShut {NoStop}%
\bibitem [{\citenamefont {Gheorghiu}\ \emph {et~al.}(2015)\citenamefont
  {Gheorghiu}, \citenamefont {Wallden},\ and\ \citenamefont
  {Kashefi}}]{Gheorghiu2015}%
  \BibitemOpen
  \bibfield  {author} {\bibinfo {author} {\bibfnamefont {A.}~\bibnamefont
  {Gheorghiu}}, \bibinfo {author} {\bibfnamefont {P.}~\bibnamefont {Wallden}},
  \ and\ \bibinfo {author} {\bibfnamefont {E.}~\bibnamefont {Kashefi}},\
  }\href@noop {} {\bibfield  {journal} {\bibinfo  {journal} {arXiv preprint
  arXiv:1512.07401}\ } (\bibinfo {year} {2015})}\BibitemShut {NoStop}%
\bibitem [{\citenamefont {Aaronson}\ and\ \citenamefont
  {Gottesman}(2004)}]{Aaronson2004}%
  \BibitemOpen
  \bibfield  {author} {\bibinfo {author} {\bibfnamefont {S.}~\bibnamefont
  {Aaronson}}\ and\ \bibinfo {author} {\bibfnamefont {D.}~\bibnamefont
  {Gottesman}},\ }\href {\doibase 10.1103/PhysRevA.70.052328} {\bibfield
  {journal} {\bibinfo  {journal} {Physical Review A}\ }\textbf {\bibinfo
  {volume} {70}},\ \bibinfo {pages} {052328} (\bibinfo {year}
  {2004})}\BibitemShut {NoStop}%
\bibitem [{\citenamefont {Howard}\ \emph {et~al.}(2014)\citenamefont {Howard},
  \citenamefont {Wallman}, \citenamefont {Veitch},\ and\ \citenamefont
  {Emerson}}]{Howard2014}%
  \BibitemOpen
  \bibfield  {author} {\bibinfo {author} {\bibfnamefont {M.}~\bibnamefont
  {Howard}}, \bibinfo {author} {\bibfnamefont {J.}~\bibnamefont {Wallman}},
  \bibinfo {author} {\bibfnamefont {V.}~\bibnamefont {Veitch}}, \ and\ \bibinfo
  {author} {\bibfnamefont {J.}~\bibnamefont {Emerson}},\ }\href@noop {}
  {\bibfield  {journal} {\bibinfo  {journal} {Nature}\ }\textbf {\bibinfo
  {volume} {510}},\ \bibinfo {pages} {351} (\bibinfo {year}
  {2014})}\BibitemShut {NoStop}%
\bibitem [{\citenamefont {Branciard}\ \emph {et~al.}(2012)\citenamefont
  {Branciard}, \citenamefont {Cavalcanti}, \citenamefont {Walborn},
  \citenamefont {Scarani},\ and\ \citenamefont {Wiseman}}]{Branciard2012}%
  \BibitemOpen
  \bibfield  {author} {\bibinfo {author} {\bibfnamefont {C.}~\bibnamefont
  {Branciard}}, \bibinfo {author} {\bibfnamefont {E.~G.}\ \bibnamefont
  {Cavalcanti}}, \bibinfo {author} {\bibfnamefont {S.~P.}\ \bibnamefont
  {Walborn}}, \bibinfo {author} {\bibfnamefont {V.}~\bibnamefont {Scarani}}, \
  and\ \bibinfo {author} {\bibfnamefont {H.~M.}\ \bibnamefont {Wiseman}},\
  }\href {\doibase 10.1103/PhysRevA.85.010301} {\bibfield  {journal} {\bibinfo
  {journal} {Phys. Rev. A}\ }\textbf {\bibinfo {volume} {85}},\ \bibinfo
  {pages} {010301} (\bibinfo {year} {2012})}\BibitemShut {NoStop}%
\bibitem [{Note3()}]{Note3}%
  \BibitemOpen
  \bibinfo {note} {In general, a control operation corresponds to the
  following. If the state of the control system is $0$, nothing is done to the
  target. On the other hand, when the state of the control is $1$, a particular
  transformation is applied to the target system.}\BibitemShut {Stop}%
\bibitem [{Note4()}]{Note4}%
  \BibitemOpen
  \bibinfo {note} {$K_i \left | G \right > = \left | G \right >$, where $\left
  | G \right >$ is the quantum graph state defined as the simultaneous
  eigenstate for all the generators $K_i = X_i \otimes _{j \in N(i)} Z_j$,
  where $j$ is in the neighbourhood of $i$ and $X$ and $Z$ are the usual pauli
  matrices.}\BibitemShut {Stop}%
\end{thebibliography}%

\end{document}